\documentclass[aps,prb,twocolumn,showpacs,superscriptaddress,groupedaddress,floatfix]{revtex4-1}
\usepackage{latexsym}
\usepackage{graphicx}
\usepackage{dcolumn}
\usepackage{bm}
\usepackage{amssymb}
\usepackage{amsmath}
\usepackage[space]{grffile}

\newcommand{\ba}{\bar{A}}

\begin{document}
\graphicspath{{C:/Users/Yonah/Google Drive/EntanglementQuench/NewFigs/}}

\newcommand{\niceref}[1] {Eq.~(\ref{#1})}
\newcommand{\fullref}[1] {Equation~(\ref{#1})}
\title{Entanglement properties of the critical quench of $O(N)$ bosons}

\date{\today}

\begin{abstract}
The entanglement properties of quenched quantum systems have been studied for a decade, however results in dimensions other than $d=1$ are generally lacking. We remedy this by investigating the entanglement properties of bosonic critical systems in $d=3$, both numerically and analytically, comparing the free and the interacting critical quench of an $O(N)$ model. We find that the evolution of the entanglement entropy for these two systems is nearly identical, as expected from the "quasi-particle" picture. However, the low-lying entanglement spectrum is controlled by the different critical exponent of the two systems, and therefore these exponents may be extracted by purely entanglement-theoretic calculations. We verify this scaling numerically.
\end{abstract}

\author{Yonah Lemonik}
\email{yl76@nyu.edu}

\author{Aditi Mitra}
\email{aditi.mitra@nyu.edu}

\affiliation{Department of Physics, New York University, 4 Washington Place, New York, New York, 10003, USA}
\maketitle

\section{Introduction\label{sec:intro}}

In a quantum quench the system is prepared in the ground state of a local Hamiltonian. The state is then evolved under a different Hamiltonian, so that the system
is in a highly excited state~\cite{Calabrese2006,Calabrese2007,PolkovnikovRMP}. The ensuing dynamics have become an object of intense study
as they exemplify strongly non-equilibrium quantum systems,
and are a natural experimental protocol for cold atom techniques~\cite{Bloch2008,Trotzky2012,Gring2012,Langen2013,Monroe14}.

A fruitful way to understand quantum quenches is through quantum information theoretic quantities such as the entanglement entropy and entanglement spectrum
(ES)~\cite{Vidal03,Vidal07,Casini07,Casini09,EntRmp10,Wenbook15,Ryu15,Greiner15}. To construct these entanglement statistics a region of the physical space is selected and then the degrees of freedom outside this region are traced out.
This transforms the wavefunction into a reduced density matrix, the properties of which define the entanglement statistics.

The power of using entanglement to characterize quenches was demonstrated in particular in $1d$ conformal systems where there is a relationship between
the growth of the entanglement entropy and universal quantities\cite{Cardy16}

\begin{equation}
S(t) = \frac{\pi v c t}{6 \tau_0} +\frac{c}{3}\log \tau_0;\qquad vt< L/2,
\end{equation}
where $L$ is the length of the selected region, $v$ the velocity, $\tau_0$ is the initial correlation length  and $c$ is the central charge of the conformal field theory. For $t > L/2$, the entanglement entropy is constant.
Many studies have built on this result, but are still mainly in $1d$~\cite{calabrese2005,calabrese2007ent,Igloi12,Huse13} or in specialized theories~\cite{hartman2013}.
Results on natural field theories are scarce as entanglement statistics are not easily calculated by standard field theoretic techniques.
Therefore it is unclear what one should expect in generic field theories in $d>1$.

In this work we consider the entanglement properties of bosonic quantum quenches in $d=3$.
We study a quench from an initial state with short ranged correlations. This may be thought of as the ground state of a free boson with a large mass. The state is then evolved under one of two different Hamiltonians. The first is a free massless Hamiltonian. We emphasize that although this is a quench between two different free Hamiltonians the behavior is non-trivial and has not yet been discussed in the literature.

The second quench is to a Hamiltonian governing the critical prethermalization state of the interacting $O(N)$ model, the latter identified in
Refs.~\onlinecite{Chiocchetta2015,Maraga2015,Chiocchetta16}.
A generic interacting theory is expected to thermalize after a quench and therefore will have all correlations described by the Gibbs ensemble.
However in some theories the thermalization time is long enough that a rich physics may appear between the non-universal regulator dependent short time scale
and the thermalization time. In the case of the $O(N)$ model the thermalization time may be made large by increasing $N$.
This prethermalization state is well described by the $N\rightarrow \infty$~
limit \cite{Sondhi2013}. The $1/N$ corrections control the thermalization behavior which is irrelevant to the short time pre-thermalization behavior. 
The $O(\infty)$ model is a non-integrable interacting system, but it is sufficiently tractable that the entanglement statistics may be directly extracted,
for relatively small system sizes.

The quenched $O(\infty)$ model has a critical transition separating coarsening and disordered regimes. This critical transition, like usual critical points, is characterized by a universal exponent $\theta$ identified in references~\onlinecite{Chiocchetta2015,Maraga2015,Chiocchetta16} called the initial slip exponent. This $\theta$ controls universal aging phenomena~\cite{Janssen1988,Huse89,Gambassi2005,Gagel2014} which appear in our system in the fluctuations of the bosonic field $\phi$,
\begin{equation}
\left\langle
	 \left\{
	 	\phi(r,t),\phi(r',t)
	 \right\}
\right\rangle
	\sim
	|r-r'|^{-2\theta  + 2 - d},
\end{equation}
when $|r-r'| \ll t$ and the velocity of excitations has been set to 1.   In the case of the free massless Hamiltonian the initial slip exponent takes the value $\theta =  0$. In the critical interacting case it takes the value $(4-d)/4$, or $\theta = 1/4$ in $d=3$.  

We emphasize that the critical exponent $\theta$ is not an equilibrium critical exponent. Rather, this exponent should actually be thought of as
a boundary critical exponent~\cite{Chiocchetta16,Gambassi2005} that governs the renormalization of the $\phi$ field at the temporal boundary $t=0$.
As it requires a temporal boundary it cannot be probed by local perturbations of the equilibrium critical state. We also note that the physics
involved is not directly related to the Kibble-Zurek mechanism~\cite{Kibble76,Zurek85}. As the quench is large and instantaneous there is no gradual
freezing out of fluctuations, and the exponent $\theta$ cannot be calculated from these considerations. Thus while the Kibble-Zurek
exponents can be derived from equilibrium critical exponents, $\theta$ in contrast is truly a new exponent arising entirely due to the quench.

The numerical calculation of the entanglement statistics of both the free and $O(\infty)$ quench proceeds by defining correlation functions on the cubic lattice. For the free case this may be trivially done. For the interacting case this is done by suitably regularizing the previously calculated long-range critical behavior in a way that is compatible with the underlying lattice. Using the fact that both states have Gaussian correlations, the entanglement statistics may then be directly calculated from the correlation functions.

\begin{figure}
\includegraphics[width = .49\columnwidth]{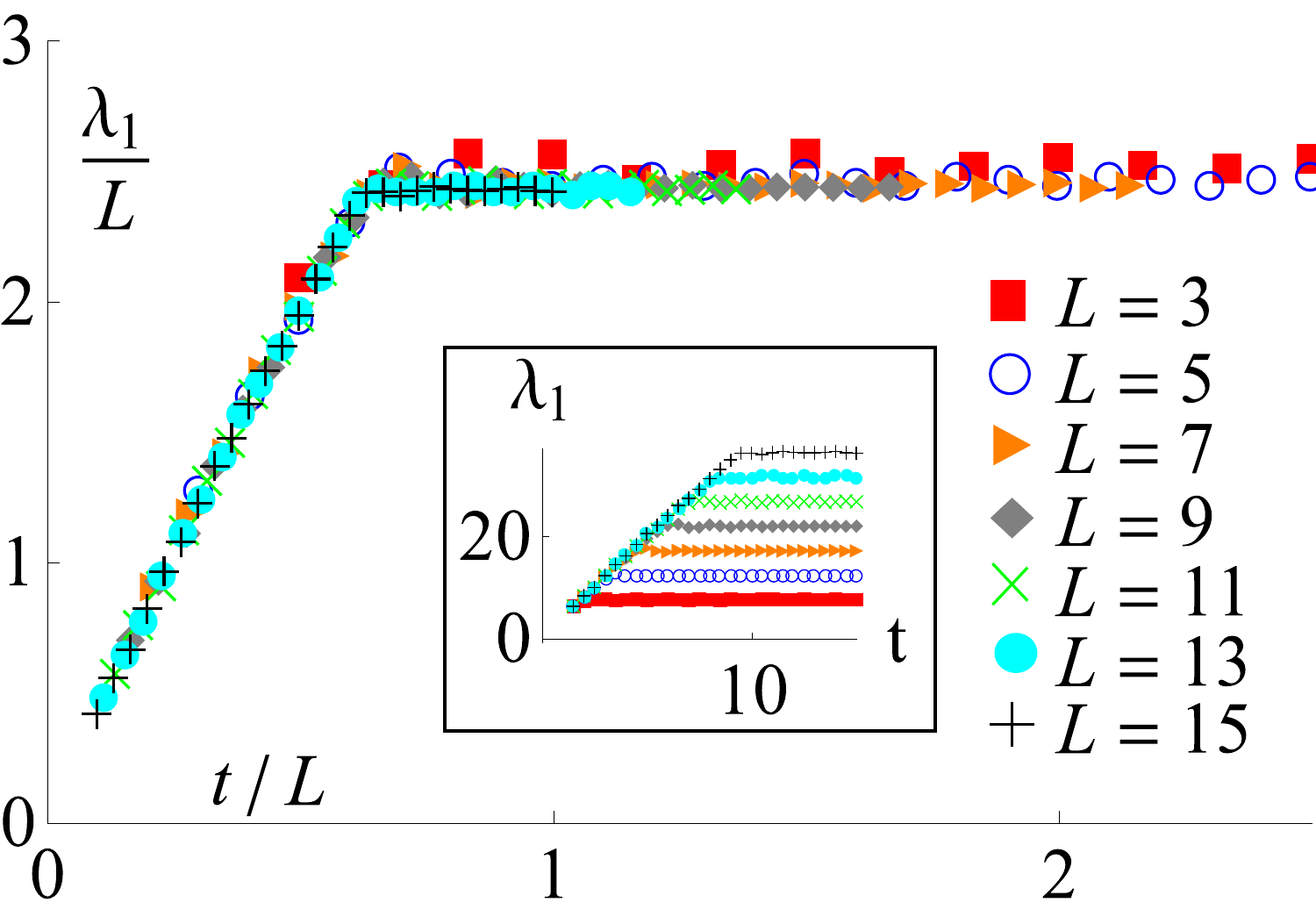}
\includegraphics[width = .49\columnwidth]{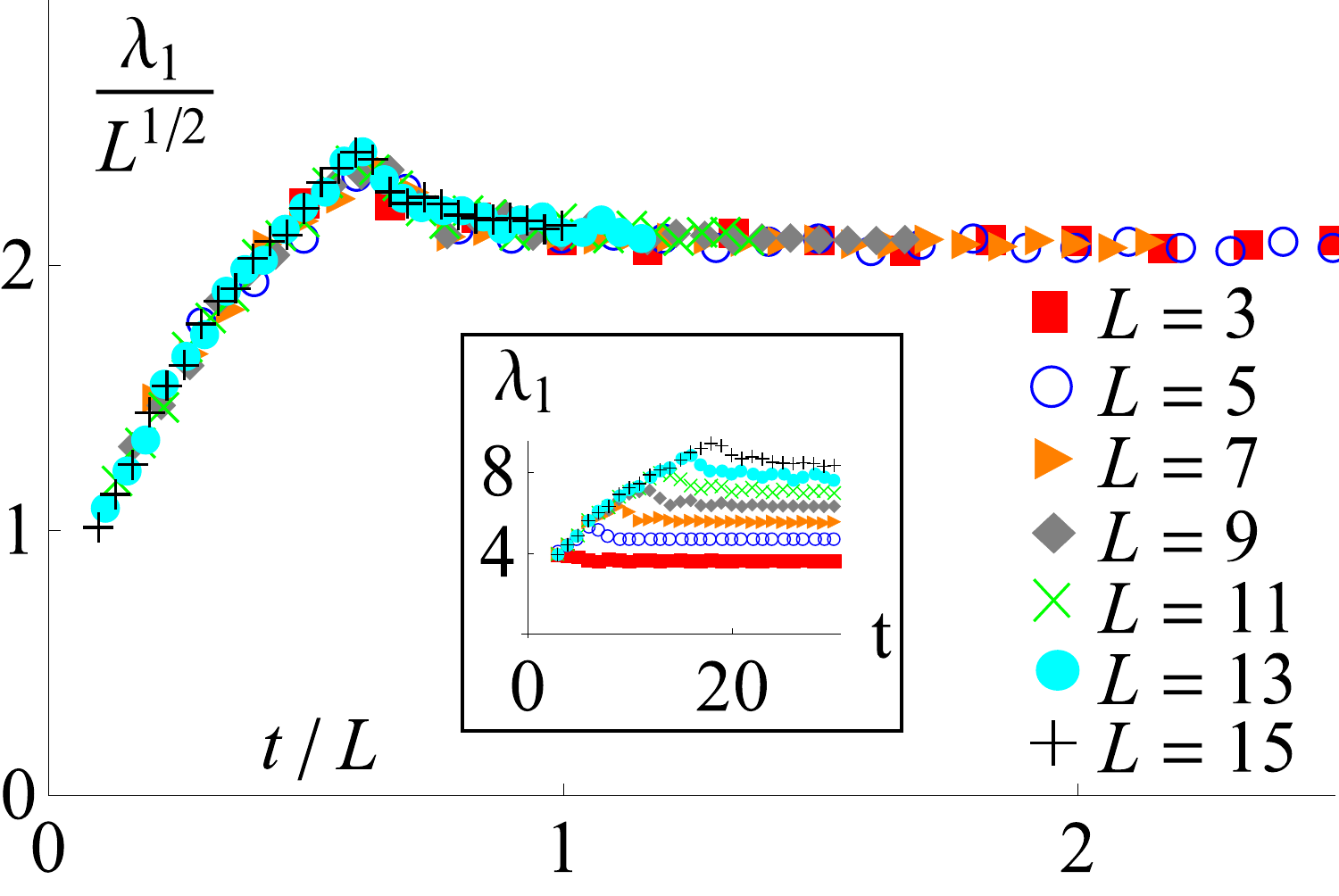}
\includegraphics[width = .49\columnwidth]{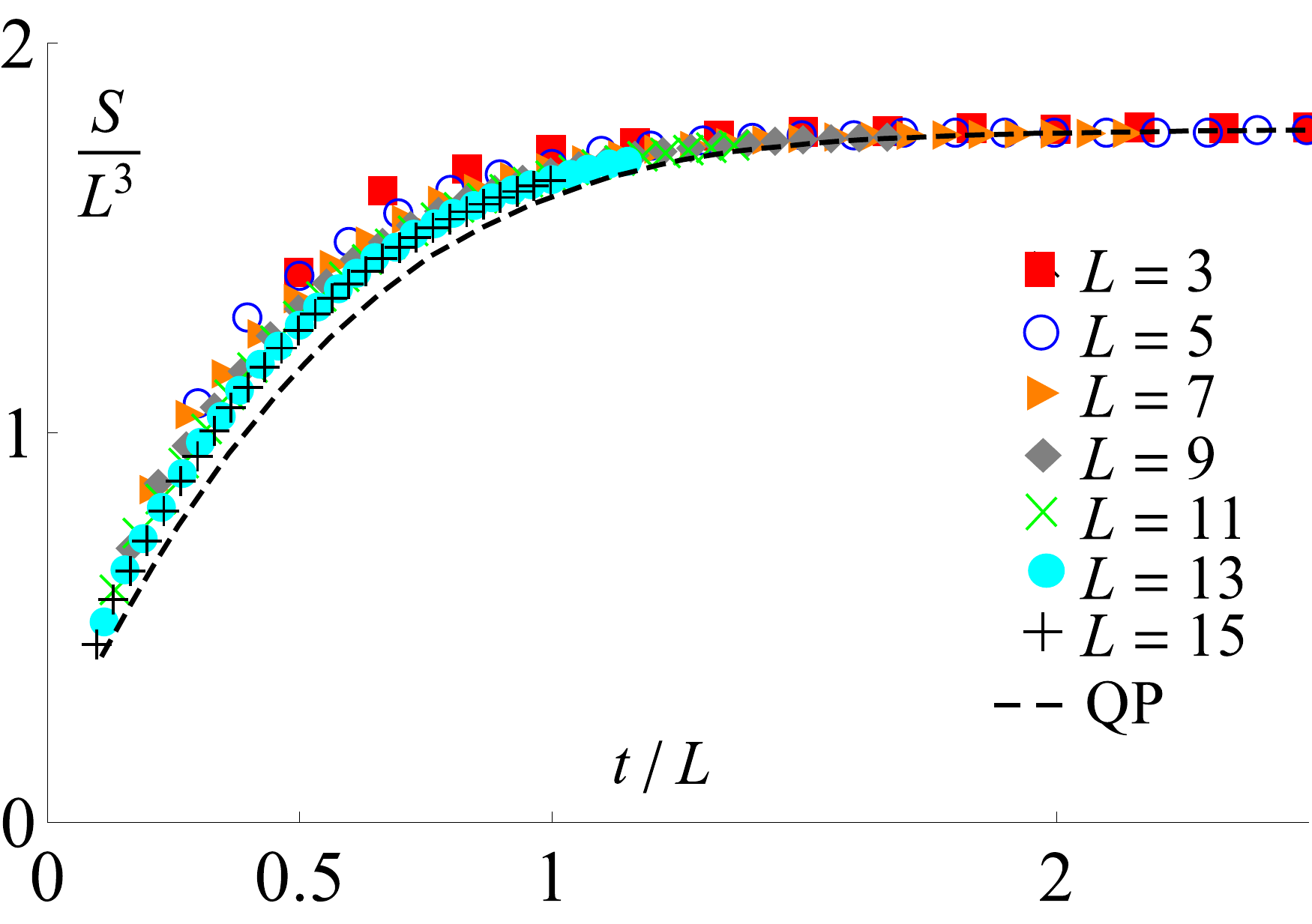}
\includegraphics[width = .49\columnwidth]{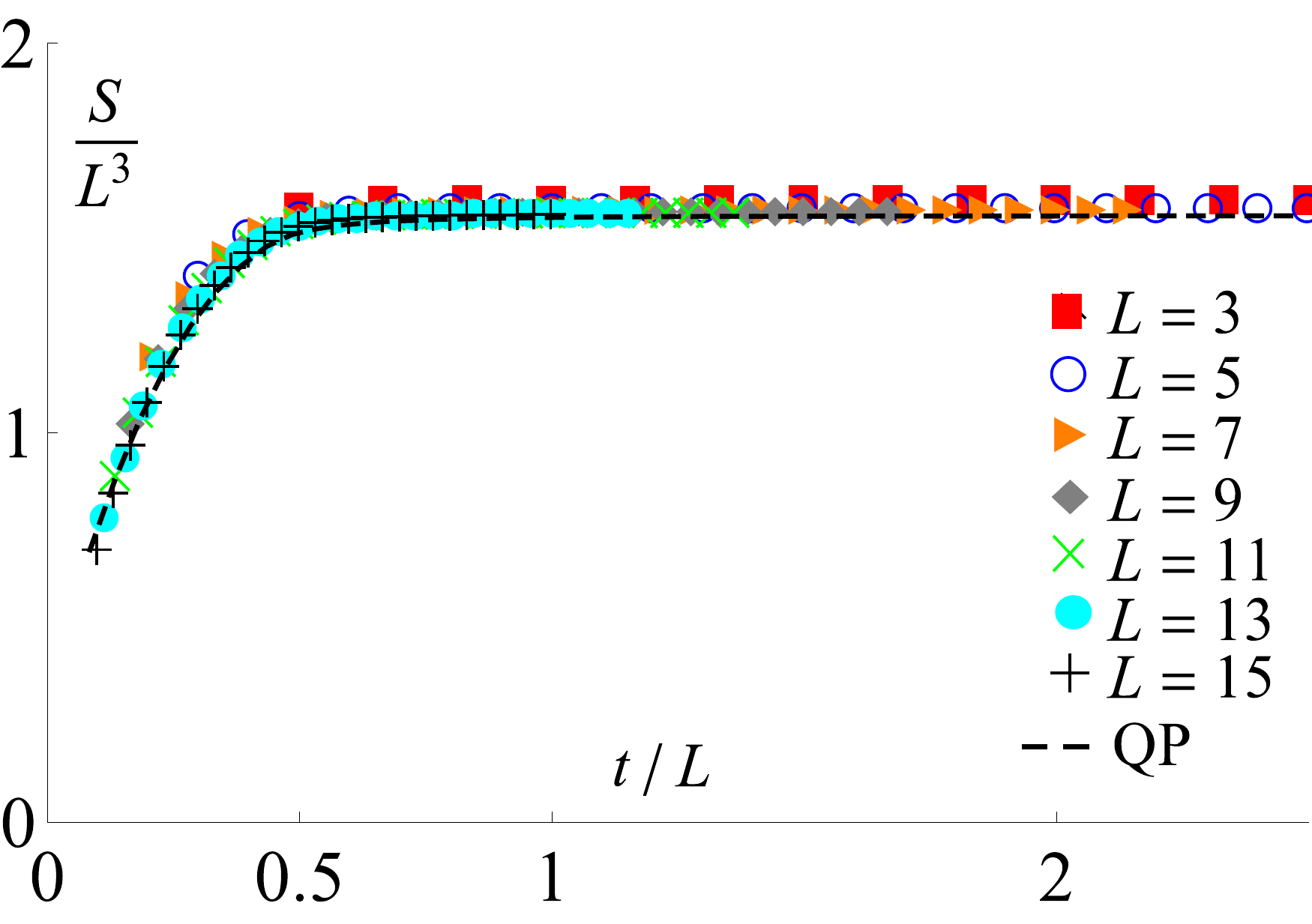}
\caption{(Color online.) Scaling of the entanglement spectrum. Top left: First eigenvalue for the free quench, scaled by exponent $1$ for different sizes $L$, inset unscaled values. Top right: First eigenvalue for the interacting quench, scaled by exponent $1/2$, inset unscaled values. Bottom left: Scaled entanglement entropy for an interacting quench with nearest neighbor dispersion for several different $L$. The dashed line shows a fit with the "quasi-particle" approximation. Bottom right: Same for the altered dispersion.\label{fig:eigscaling}}
\end{figure}

Combining this numerical computation and analytic arguments we arrive at two principal conclusions. First, the time evolution of the entanglement entropy is well described by the ``quasi-particle'' picture~\cite{calabrese2007ent} and controlled by non-universal properties of the dispersion, Fig.~\ref{fig:eigscaling}, lower panels.
Second, the scaling of a part of the eigenvalues of the ES is universal, Fig~\ref{fig:eigscaling}, upper panels. 
That is the largest eigenvalues $\lambda$ of the spectrum obey a scaling relation,
\begin{equation}
\lambda(L,t) = L^{-2\theta + 1} W(t/L),
\end{equation}
where $L$ is the linear size of the entanglement region and $W$ is some unknown scaling function. The existence of such a scaling relation opens up the possibility of determining non-equilibrium critical exponents in systems where they are unknown.

The ability to cleanly separate universal and non-universal physics stems from the clear scaling behavior of the
non-equilibrium phase transition. Since an effective temperature generated by a quench causes such phase transitions to only exist
in $d > 2$, our results are a significant advance over previous studies in $d=1$.

The remainder of the paper is structured as follows. In Sec.~\ref{sec:quenches}, we lay out the physical models under consideration and discuss their correlation functions. In Sec.~\ref{sec:entangle}, we discuss how the ES is calculated and present numerical results for it. In Sec.~\ref{sec:analytic}, we use analytic arguments to derive the structure of the ES. We conclude in Sec.~\ref{sec:conclusion}.

\section{Critical Quenches \label{sec:quenches}}
\subsection{System}
We consider a system of bosons on a three dimensional cubic lattice, described by canonically commuting operators $\phi^a_i$, $\pi^b_i$, $[\phi^a_i,\pi^b_j] = i\delta_{ab}\delta_{ij}$ , where $ij$ indicate lattice positions and $ab$ some additional flavor index that takes
$N$ values. At time $t=0$ the system is in the ground state of the Hamiltonian
\begin{equation}
H_{\rm init}  \equiv \frac{1}{2}\sum_{ai}\left[ \phi^a_i t_{ij}\phi^a_j + \omega_0^2 \left(\phi^a_i\right)^2+  \left(\pi^a_i\right)^2\right].
\end{equation}
Here $t_{ij}$ is a translationally invariant hopping matrix.  The matrix $t_{ij}$ determines  the dispersion $\varepsilon_k^2 = \sum_j \exp\left( ik\cdot r_j\right) t_{ij}$. We take  $\varepsilon_k  = 0$ only at $k=0$ and
near $k \sim 0$ we assume ${\varepsilon}(k) \sim c |k|$, i.e., the boson is massless when $\omega_0 = 0$. The constant $c$ is the speed of low energy quasi-particles. We rescale $t$ so that $c = 1$. These conditions determine the  long distance properties of the dispersion.

In the initial state we take the large mass limit $\omega_0^2 \gg t_{ij}$
where the operators have local correlations,
\begin{equation}
\langle \phi^a_i \phi^b_j \rangle  = \frac{\omega_0}{2} \delta_{ij}\delta_{ab};
\quad
\langle \pi^a_i \pi^b_j \rangle =  \frac{1}{2\omega_0} \delta_{ij} \delta_{ab};
\quad
\langle \pi^a_i \phi^b_j \rangle  = 0,
\label{eq:initcon}
\end{equation}
with higher correlations determined by Wick's theorem. For $t >0$ the system is evolved under a local Hamiltonian.

We consider two possibilities. First the free massless Hamiltonian
\begin{equation}
H_{\rm free} \equiv \frac{1}{2}\sum_{ai}\left[ \phi^a_i t_{ij}\phi^a_j+  \left(\pi^a_i\right)^2\right].
\end{equation}
In order for this to be massless we must set the appropriate conditions on $t_{ij}$.

The second quench is to the interacting Hamiltonian
\begin{equation}
H_{\rm int} \equiv H_{\rm free} + \frac{m_0^2}{2}\sum _{ia}
\left(\phi^a_{i}\right)^2 + \frac{u}{4!N}\sum_{iab}\left(\phi^a_i\right)^2\left(\phi^b_i\right)^2.
\end{equation}

In both cases the initial state corresponds to a highly excited state, with an extensive amount of energy. As we shall see, the correlations  of $H_{\rm free}$ possess a scale free character, coming from the lack of a low energy scale. Similarly, it was shown that by tuning $m_0$ a similar scale free structure is obtained for the correlations under
$H_{\rm int}$~\cite{Sciolla2013,Gambassi11,Sondhi2013,Smacchia2015,Chiocchetta2015,Maraga2015}.

\begin{figure}
\includegraphics[width = .5\columnwidth]{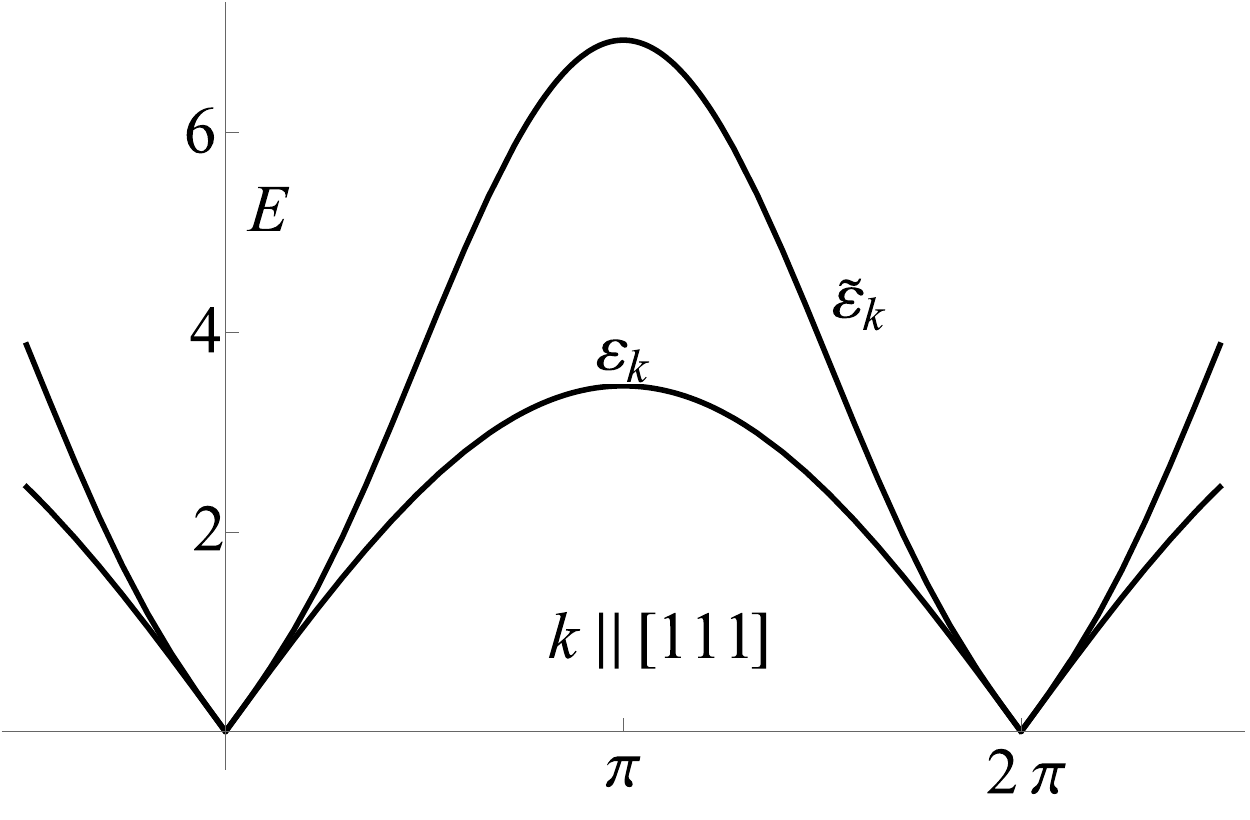}\includegraphics[width = .5\columnwidth]{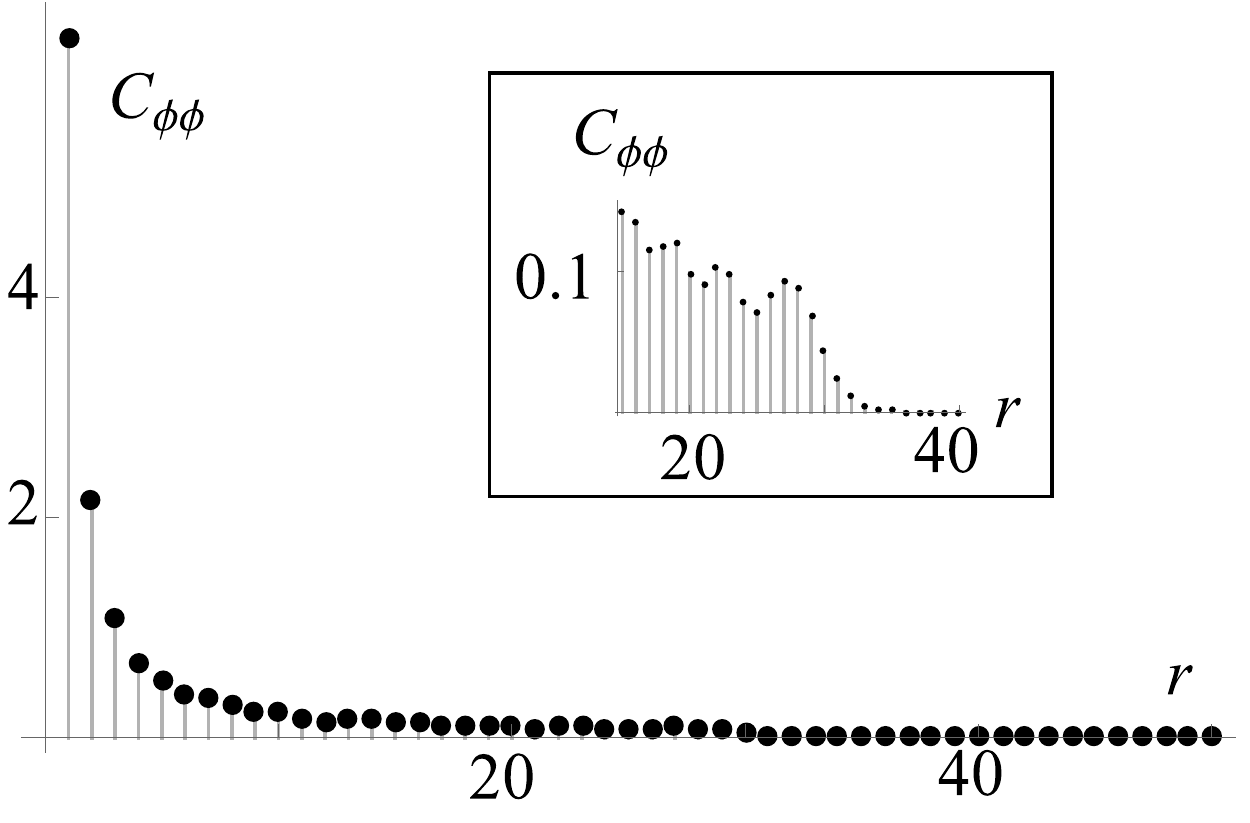}
\includegraphics[width = .5\columnwidth]{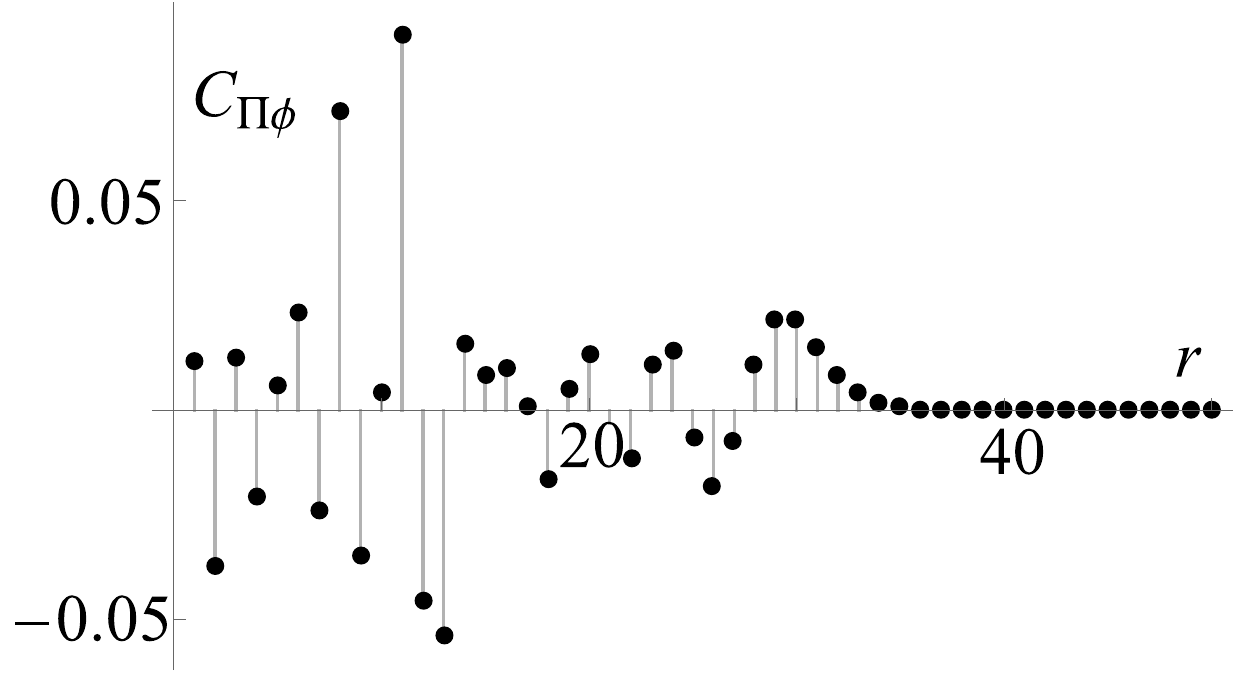}\includegraphics[width = .5\columnwidth]{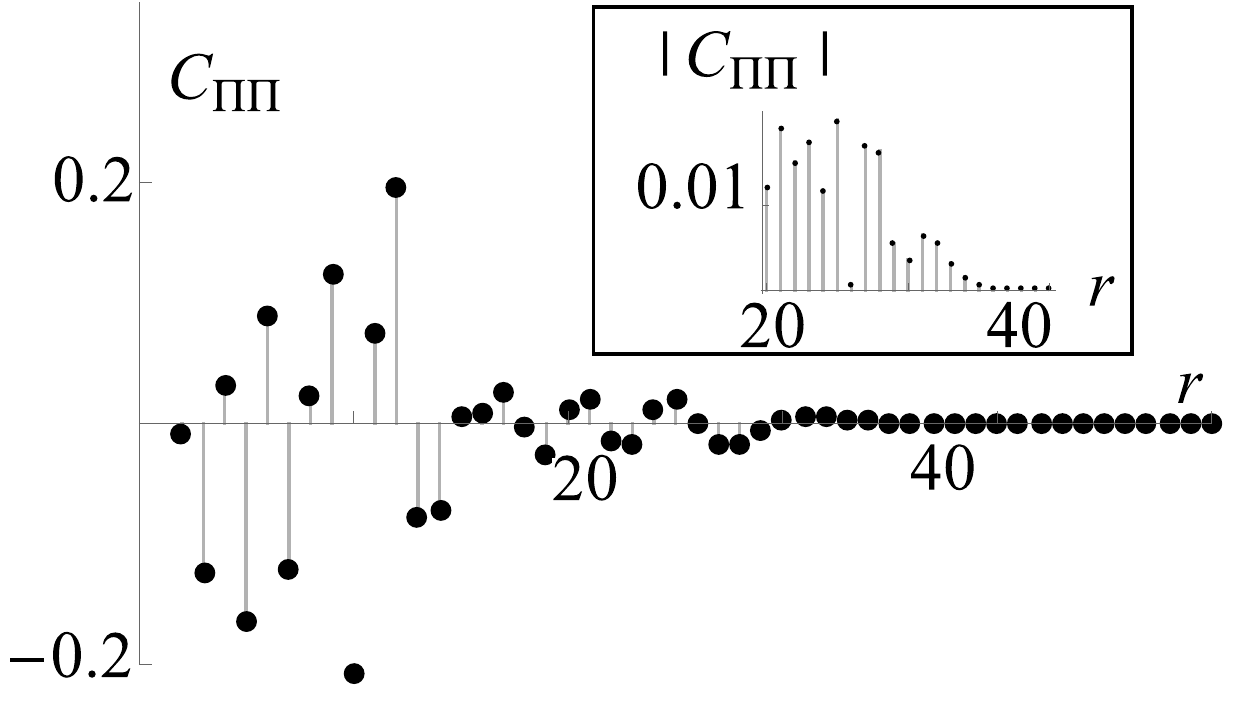}
\caption{(Color online.) Top left: Nearest neighbor dispersion $\varepsilon(k)$  and the alternate dispersion $\tilde{\varepsilon}(k)$, along the $[111]$ direction through the BZ. Remaining panels: The three correlations as a function of position at $t=15$. Top right inset
and bottom right inset is the zoom of light cone shoulder.
At $r=0$, $C_{\pi\pi} \gg 1$ (not shown), a remnant of the initial condition given in ~\niceref{eq:initcon}.\label{fig:Correlators}}
\end{figure}

\subsection{Correlation Functions} We study the $N\rightarrow \infty$ limit, where self-consistent Hartree-Fock is exact. In this case, Wick's theorem holds and all the entanglement quantities may be extracted from the Keldysh equal-time correlators
\begin{equation}
C_{\mathcal{OO}'}\left(r-r';t\right)
	\equiv
\langle
	\left\{
		\mathcal{O}\!\left(r,t\right)
		,
		\mathcal{O}'\!\left(r',t\right)
	\right\}
\rangle/2
\end{equation}
where $\mathcal{O},\,\mathcal{O}'$ are $\pi$ or $\phi$. There are three distinct correlators, which may be written as,
\begin{subequations}
\begin{align}
C_{\phi\phi}\left(r;t\right) &= \int_{BZ} \frac{d^3 k}{(2\pi)^3}\, |f_k(t)|^2 e^{i\vec{k}\cdot \vec{r}}\\
C_{\Pi\phi}\left(r;t\right) &= \int_{BZ} \frac{d^3 k}{(2\pi)^3}\, {\rm Re}\left[f_k^*(t)\partial_t f_k(t)\right] e^{i\vec{k}\cdot \vec{r}}\\
C_{\Pi\Pi}\left(r;t\right) &= \int_{BZ} \frac{d^3 k}{(2\pi)^3}\, |\partial_t f_k(t)|^2 e^{i\vec{k}\cdot \vec{r}},
\end{align}
\label{eq:defC}
\end{subequations}
where the commutation relation imposes
\begin{equation}
2 \text{Im}[\dot{f}_k f^*_k] = 1. \label{eq:uncertainty}
\end{equation}
For the free quench $f_k(t)$ is
\begin{equation}
f_k(t) =
	i\sqrt{\frac{\omega_0}{2}}\frac{\sin(\varepsilon_k t)}{\varepsilon_k } +
	\frac{1}{\sqrt{ 2\omega_0} }\cos(\varepsilon_k t)
\end{equation}
In the case of the interacting system, Ref.~\onlinecite{Maraga2015} analyzed the behavior of $f_k$ at small $k$. They found that for some critical value of $m_0$,  $f_k(t)$ for $t$ greater than the lattice spacing is given by
\begin{equation}
f_k(t) \!=
	\!\! \sqrt{\frac{t}{2}}
	 \left[ 		
		 \kappa_+
		 \left(
		 	\frac{\varepsilon_k}{\omega_0}
		 \right) ^{\!-\alpha}\!\!\!
		 j_\alpha\left(\varepsilon_k t\right)
		 +
		 \kappa_-
	     \left(
		 	\frac{\varepsilon_k}{\omega_0}
	  	 \right)^{\!\alpha}\!\!\!
	  	 j_{-\alpha}\left(\varepsilon_k t\right)
	 \right] ,\label{eq:defIntf}
\end{equation}
where since their expressions were derived for long wavelengths, $\varepsilon_k =|k|$,
with $j_\alpha(x)$ the Bessel function. $\kappa_{\pm}$ are non-universal constants that obey
$\text{Im}[\kappa_+ \kappa_-^*] = -\pi/(2\sin\left(\pi\alpha\right))$,
and depend on the details of the dispersion and initial condition.  The exponent
\begin{eqnarray}
\alpha \equiv -\theta + 1/2 =(d-2)/4;
\end{eqnarray}
in dimension $d>2$. This exponent controls aging effects when the system is probed at two different times~\cite{Janssen1988,Huse89,Chiocchetta2015}. At the upper critical dimension of $d=4$, $\alpha = 1/2$ and $f$ reduces to the free case.

Although we are studying unitary quantum dynamics, an effective temperature generated by the quantum quench gives the same upper critical
dimension as for the classical theory. For finite $N$, inelastic scattering will cause the system to thermalize after a time $t^*\sim {\cal O}\left(N\right)$
resulting in diffusive rather than ballistic propagation of quasi-particles. For times $t < t^*$ the system is in a prethermal regime which is qualitatively
similar~\cite{Chiocchetta2015,Chiocchetta16} to the $N\rightarrow\infty$ limit.

The correlation functions given by \niceref{eq:defIntf} cannot be directly applied to the entanglement entropy,
since they are only valid at small $k$ and therefore  must be regulated.
As we shall see in Sec. IIIA, it is important to choose a regulator that maintains the commutation relations and the uncertainty principle, as otherwise the entanglement spectrum
will be unphysical and lead to imaginary entropies and other pathology. This means that various common choices of regulator, such as adding an exponential decay to
Eq.~\eqref{eq:defC} cannot be used.

Instead, we regulate the expression by taking the form of Eq.~\eqref{eq:defIntf} and substitute for $\varepsilon_k$ a function periodic in the Brilloiun zone,
with the appropriate low energy properties discussed in Sec. IIA. In this way the commutation relation Eq.~(\ref{eq:uncertainty}) is exactly maintained,
and the real space correlation functions can be calculated by Fourier transform. We emphasize that unlike in the free case, where this substitution is exact,
in the interacting case this is simply a choice of regulator. Nonetheless this should be irrelevant to the extraction of universal behavior.

We now consider the real space behavior of the correlation functions. If $r$, $t$ \emph{and} $|r-2t|$ are large compared to the lattice scale then  Eq.~\eqref{eq:defC} may be evaluated by taking $|k|$ small, substituting $\varepsilon_k \rightarrow |k|$ and dropping all oscillatory components, giving,
\begin{align}
C_{\phi\phi}\left(r \ll 2t\right)&  \sim r^{2\alpha + 1 -d},\\
C_{\phi\phi}\left(0< 2t-r \ll t\right) &\sim  r^{\frac{1-d}{2}} \left(2t-r\right)^{2\alpha + \frac{1}{2} - \frac{d}{2}},\\
C_{\mathcal{OO'}}\left(2t < r\right)& \sim 0.
\end{align}
The above expressions can be obtained not only from solving the $N=\infty$ problem~\cite{Maraga2015}, but also from performing a dimensional expansion~\cite{Chiocchetta16}.

The singularity as $r\rightarrow 2t$ and the suppression of the correlation function when $r > 2t$ is given the following interpretation~\cite{Calabrese2006}. At $t=0$, an extensive amount of energy is injected into the system. This may be thought of as quasi-particles being emitted isotropically from all points. For $t>0$ these quasi-particles propagate ballistically with velocity $\sim 1$.  As the initial state is only locally correlated, these quasi-particles are correlated only with other quasi-particles emitted from the same point. When $|r-r'|>2t$, no correlated pair of quasi-particles has reached the points $r$ and $r'$, so the correlations are zero. When $|r-r'|= 2t$, correlated pairs emitted from the midway point between $r$ and $r'$ arrive, leading to a singular feature.
Similarly for  $C_{\Pi\phi}$ and $C_{\Pi\Pi}$, we find,
\begin{align}
C_{\Pi\phi}\left(r \ll 2t\right) &\sim r^{2\alpha+1 -d}t^{-1}\\
C_{\Pi\phi}\left(0 < 2t-r \ll t\right) &\sim  r^{\frac{1-d}{2}} (2t-r)^{2\alpha - \frac{1}{2} - \frac{d}{2}}\\
C_{\Pi\Pi}\left(r \ll 2t\right) &\sim r^{2\alpha -1 -d}\\
C_{\Pi\Pi}\left(0 < 2t-r \ll t\right) &\sim  r^{\frac{1-d}{2}} (2t-r)^{2\alpha - \frac{3}{2} - \frac{d}{2}}
\end{align}
Figure~\ref{fig:Correlators} highlights this behavior.

The existence of power laws is indicative of the scale free nature of the quench. Further, we may obtain a scaling form for $C_{\mathcal{OO}'}$  for general $r$ and $t$.  Consider the limit where $t \rightarrow \infty$ but $k t$ remains finite, $f_k(t)$ behaves as:
\begin{equation}
f_k(t) = t^{\alpha+1/2}h(kt) + \text{sub-leading as }t \rightarrow \infty.
\end{equation}
The neglect of the subleading terms is non-trivial since it violates the commutation relation, \niceref{eq:uncertainty}. Neglecting this issue, inserting the expansion into Eqs.~\eqref{eq:defC} and rescaling the integration variable $k \rightarrow q/t$, we obtain that at times large compared to the non-universal scales the correlation functions are
\begin{subequations}
\begin{align}
C_{\phi\phi} &= t^{2\alpha + 1 -d} g_{\phi\phi}\left(r/t\right) \\
C_{\phi\Pi} &= t^{2\alpha -d} g_{\phi\Pi}\left(r/t\right) \\
C_{\Pi\Pi} &= t^{2\alpha - 1 -d} g_{\Pi\Pi}\left(r/t\right).
\end{align}
\end{subequations}
On first sight, the different correlation functions scale with different exponents.
However, this is not a ``coordinate-free'' statement as we are free to make a symplectic transformation of $\pi$ and $\phi$. In particular, if
we make the symplectic transformation $\phi\rightarrow \phi/\sqrt{t}$, $\Pi\rightarrow \Pi \sqrt{t}$, all correlations scale as
\begin{equation}
C_{\mathcal{OO}'}(r,t) = t^{2\alpha-d}g_{\mathcal{OO}'}(r/t).\label{eq:scalingC}
\end{equation}

It is important to note that the relevance of Eq.~(\ref{eq:scalingC}) to the entanglement statistics is a priori unclear for the quenched system. The scaling form only holds at long ranges, whereas entanglement statistics for an arbitrary quantum state are not naturally divided into long and short range parts. As we shall see, it is only a small subset of the entanglement that actually obeys the universal scaling.

\section{Calculation of Entanglement Statistics\label{sec:entangle}}
\subsection{Entanglement Spectrum of Gaussian Bosons}

The essential idea of entanglement statistics is that in lieu of studying the state of the system $|\Psi\rangle$ directly, we instead study a reduced density matrix constructed by tracing over part of our Hilbert space. In our case, the Hilbert space $\mathcal{H}_\text{tot}$ is composed of tensor products of countably infinite Hilbert spaces,
\begin{equation}
\mathcal{H}_\text{tot} = \otimes_{ia} \mathcal{H}_i^a,
\end{equation}
where as before $i$ indicates a lattice site and $a$ the flavor index, one for each of the harmonic oscillators present in the system. We may rewrite our state $|\Psi\rangle$ with respect to this partition of the Hilbert space. Enumerating the a basis of $\mathcal{H}_i^a$ as $|\varphi_\alpha\rangle_i^a$ we write,
\begin{equation}
|\Psi\rangle = \sum_{\alpha\beta\ldots} \psi^{\alpha \beta\ldots} |\varphi_\alpha\rangle_1^1 \otimes |\varphi_\beta\rangle_2^1\otimes\cdots.
\end{equation}

Now we select some region of the lattice $R$ according to which the Hilbert space factorizes into two parts
\begin{equation}
\mathcal{H}_\text{tot} = \mathcal{H}_\text{in}\otimes\mathcal{H}_\text{out}.
\end{equation}
Then we perform the partial trace over the Hilbert space $\mathcal{H}_\text{out}$. This maps the state $|\Psi\rangle$ to a reduced density matrix $\rho_R$ given by the partial trace,
\begin{equation}
\rho_R = {\rm Tr}_\text{out}\left[|\Psi\rangle\langle\Psi|\right].
\end{equation}

A statistic of $\rho_R$ is the entanglement spectrum, which is simply the spectrum of the operator $\rho_R$.  This contains the information which is invariant under  arbitrary unitary transformations of $\mathcal{H}_\text{in}$, i.e., all of the information which knows only about the entanglement of region $R$ and its complement but not about the internal structure of $R$ (or its complement).

We wish to calculate the entanglement spectrum of the density matrix $\rho_R(t)$ given by the time evolution of the quenched state, with respect to some region $R$. We largely follow Ref.~\onlinecite{Eisler2009}.
In the case under consideration, a simplification is that at all times we have Gaussian correlations that are determined by Wick's theorem. The fact that Wick's theorem determines all correlation is still true if we only consider correlation functions between operators in $R$.
Therefore, Wick's theorem is valid for expectations with respect to $\rho_R$.

Any density matrix for which Wick's theorem applies can be written as
\begin{equation}
\rho_R = \exp\left(-\frac{1}{2}\sum_{ij} X_i M_{ij} X_j \right),\label{eq:defM}
\end{equation}
where we have introduced the generalized coordinates $\vec{X} = \left(\vec{\phi},\, \vec{\Pi}\right)$, which combines the position index and the operators $\phi$ and $\Pi$.  Let us also introduce the total correlation matrix $\bf{C}$, whose elements are given by
\begin{equation}
{\bf C}_{ij}(t) = \langle \left\{X_i(t), X_j(t)\right\}\rangle/2.\label{eq:defbfC}
\end{equation}
The matrix elements of ${\bf C}$ are  the correlation functions already discussed.

To proceed we want to choose new coordinates $Y$ that diagonalize ${\bf C}$, and therefore $M$. 
The slight complication is that this is a bosonic system, and our new coordinates have to maintain the canonical commutation relations
\begin{equation}
[\Pi_i,\phi_j] = -i\delta_{ij},\qquad[\Pi_i,\Pi_j] = [\phi_i,\phi_j] =0,
\end{equation}
or in terms of the generalized coordinates
\begin{equation}
[X_i,X_j] = \openone\otimes \begin{pmatrix} 0  & i\\ -i & 0 \end{pmatrix} \equiv \Omega.
\end{equation}
If we make new coordinates as a linear combination by $Y_i = U_{ij} X_j$, then $Y_i$ maintain the commutation relations iff
\begin{equation}
U \Omega U^T = \Omega,
\end{equation}
that is, $U$ must be a symplectic matrix and we are seeking a symplectic diagonalization of $M$ and $C$.

Assume we have found coordinates $\vec{Y} =  \left(\vec{\phi'}, \vec{\Pi'}\right)$ such that
\begin{align}
\rho_R &= \exp\left[-\sum_i\frac{1}{2}\omega_i\left(\Pi'^2_i + Q'^2_i\right)\right] \\
	   &= \exp\left[-\frac{1}{2} Y\left( \text{diag}\left\{\omega_i\right\}\otimes\begin{pmatrix}1 & 0 \\0 &1\end{pmatrix}\right) Y\right].\label{eq:diagrho}
\end{align}
Thus $Y_i$ are the coordinates in which the ``Hamiltonian'' is that of uncoupled harmonic oscillators with frequency $\omega_i$. 
Because the system is linear these coordinates are linearly related $Y = UX$.

Taking the Keldysh correlator of $Y_i$, we get
\begin{equation}
 \frac{\text{Tr}\left[\rho_R \left\{\vec{Y},\vec{Y} \right\} /2 \right]}{\text{Tr}\left[\rho_R\right]}
  = \frac{1}{2}\text{diag}\left\{\coth(\frac{\omega_i}{2})\right\}\otimes\begin{pmatrix}1 & 0 \\0 &1\end{pmatrix}.
\end{equation}
On the other hand, if we substitute $Y = UX$ into Eq.~\eqref{eq:defbfC} we obtain that 
\begin{equation}
\frac{\text{Tr}\left[\rho_R \left\{\vec{Y},\vec{Y} \right\} /2 \right]}{\text{Tr}\left[\rho_R\right]}
  =  U^T {\bf C} U.
\end{equation}
Therefore 
\begin{equation}
 U^T {\bf C} U = \frac{1}{2}\text{diag}\left\{\coth(\frac{\omega_i}{2})\right\}\otimes\begin{pmatrix}1 & 0 \\0 &1\end{pmatrix}.
 \end{equation}
Similarly substituting $Y = UX$ into equation Eq.~\eqref{eq:defM} and comparing with Eq.~\eqref{eq:diagrho} gives
\begin{equation}
U^T M U = \left( \text{diag}\left\{\omega_i\right\}\otimes\begin{pmatrix}1 & 0 \\0 &1\end{pmatrix}\right).\label{eq:diagM}
\end{equation}
The matrices ${\bf C}$ and $M$ are therefore both diagonalized by basis given by the coordinates $Y_i$.

Now if we left multiply both sides of Eq.~\eqref{eq:diagM} by $\Omega$  we get
\begin{align}
\Omega\left( \text{diag}\left\{\omega_i\right\}\otimes\begin{pmatrix}1 & 0 \\0 &1\end{pmatrix}\right) &= \Omega U^T M U,\\
 \text{diag}\left\{\omega_i\right\}\otimes\begin{pmatrix}0 & i \\-i &0\end{pmatrix} &=  U^{-1}\Omega M U.
\end{align}
Therefore the matrices $\Omega M$ and $\text{diag}\left\{\omega_i\right\}\otimes\begin{pmatrix}0 & i \\-i &0\end{pmatrix}$ are similar and have the same eigenvalues. 
By inspection, the eigenvalues of the second matrix are $\pm \omega_i$.  Similarly the matrix 
\begin{equation}
\tilde{C} \equiv \Omega {\bf C}, \label{eq:defCtilde}
\end{equation}
has the same eigenvalues as the matrix $\frac{1}{2}\text{diag}\left\{\text{coth}\,\omega_i\right\}\otimes\begin{pmatrix}0 & i \\-i &0\end{pmatrix}$.

Therefore to find the entanglement spectrum we must construct the matrix $\tilde{C}$ and find its eigenvalues. These should be purely real and come in pairs $\pm\lambda_i$. These are related to the $\omega_i$ by
\begin{equation}
\frac{1}{2}\coth\frac{\omega_i}{2} = |\lambda_i|.
\end{equation}
Note that it is the $\lambda_i$ that we refer to as the ``entanglement eigenvalues'' and that are plotted throughout this paper.

We now relate the oscillator frequencies $\omega_i$ to the entropy of the density matrix $\rho_R$. This is precisely the same calculation as the entropy of a collection of oscillators at finite temperature,
\begin{equation}
S_R = \sum_i \left[ \frac{\omega_i e^{-\omega_i}}{1-e^{-\omega_i} } -\log\left(1- e^{-\omega_i}\right)\right].
\end{equation}
Lastly substituting in the relationship between the $\omega_i$ and the symplectic eigenvalues we obtain,
\begin{align}
S_R = \sum_i \bigg[& \left(\lambda_i + \frac{1}{2}\right)\log\left(\lambda_i + \frac{1}{2}\right)\nonumber\\
&-
\left(\lambda_i -\frac{1}{2}\right)\log\left(\lambda_i-\frac{1}{2}\right)\bigg].
\end{align}

\subsection{Numerical Calculation}

For the numerical calculation of the entanglement spectrum, we partition the lattice into a cubic region and its complement, and compute the full ES with respect to this partition. The first step is the computation of the integrals Eq.~\eqref{eq:defC}. To do this, we take our system to be periodic with length $L_{\rm tot} \gg t$. The integrals over momenta are then converted to discrete sums over momenta, which may be computed rapidly by fast Fourier transform. Since the light-cone only spreads a distance $2t$, the correlation functions do not know that the system is finite, and the results are insensitive to the value $L_{\rm tot}$.

In the second step, the sub-matrices of $C_{\mathcal{OO}'}\left(r-r';t\right)$ are constructed, where $r$ and $r'$ are restricted to lattice points in the cubic region. The total correlation matrix
is formed and diagonalized.
The ES was calculated as a function of time for cubic regions up to side-length $L=15$. We report data for both free and interacting quenches. To test sensitivity to non-universal factors we considered two dispersions: the nearest neighbor (n.n.) dispersion $\varepsilon_k^2 = 6-2\sum_{i=1,2,3}\cos k_i$ and the ``altered'' dispersion, $\tilde{\varepsilon}_k^2 = \varepsilon_k^2 + \varepsilon_k^4$. We also consider varying the energy injected in the quench by changing the parameter $\omega_0$, showing a ``low''  and ``high'' energy quench with energy per lattice site $12.5$ and $50$ respectively, for both the free and interacting systems.

\section{Behavior of the Entanglement Spectrum\label{sec:analytic}}
In this section we will discuss the structure of the entanglement spectrum. First, in Sec. IV A, we summarize the behavior of the numerically calculated spectrum. Then, in Sec. IV B, we show how the entanglement entropy may be calculated from the ``quasi-particle'' approximation. In Sec. IV C, we show how the quasi-particle approximation does not apply to the largest eigenvalues in the entanglement spectrum. These are instead governed by universal scaling laws. Lastly, in Sec. IV D, we discuss the very short time behavior when $t\sim 1$.

\subsection{Summary of numerical results}

\begin{figure}
\includegraphics[width = .49\columnwidth]{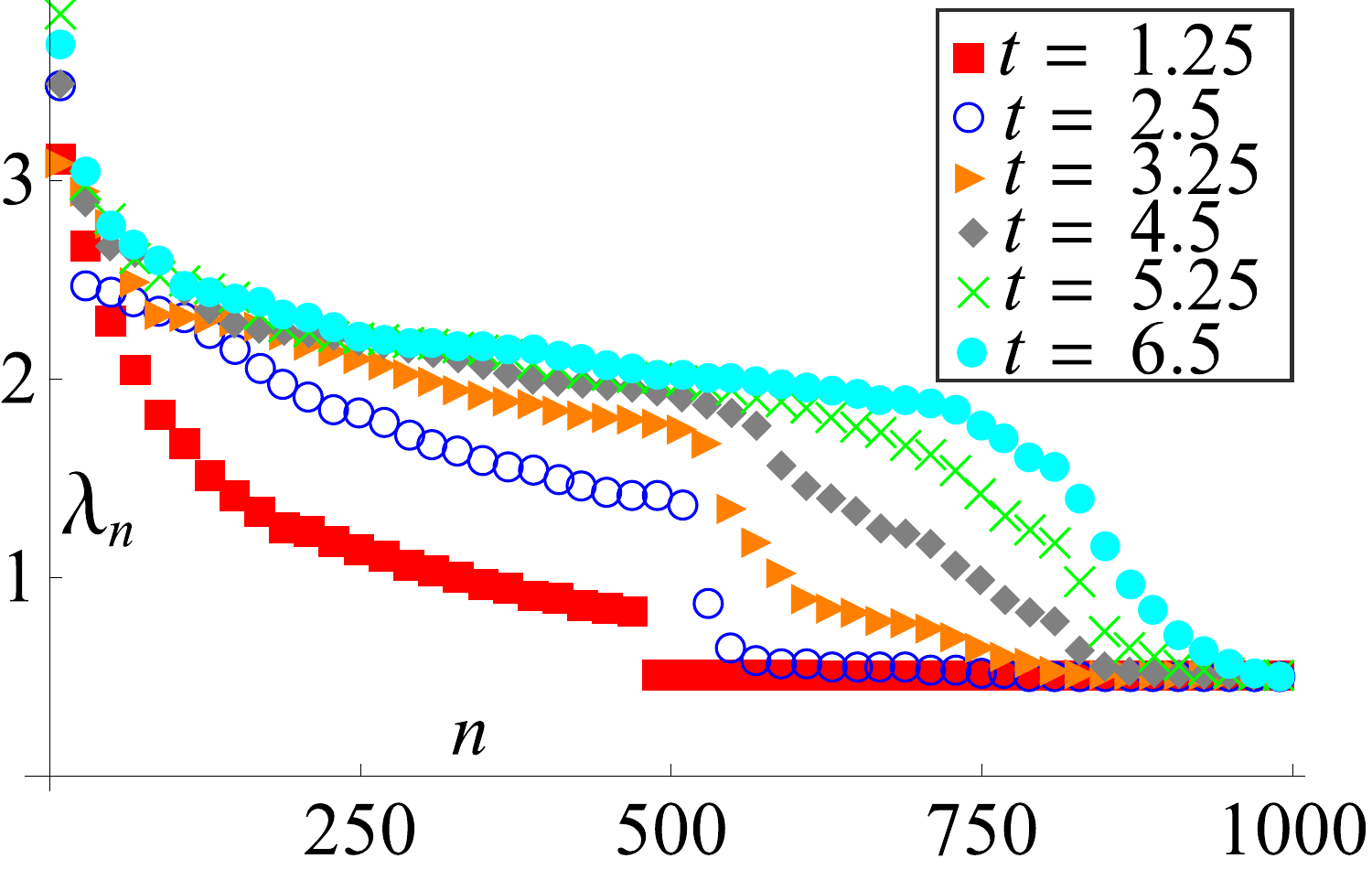}
\includegraphics[width = .49\columnwidth]{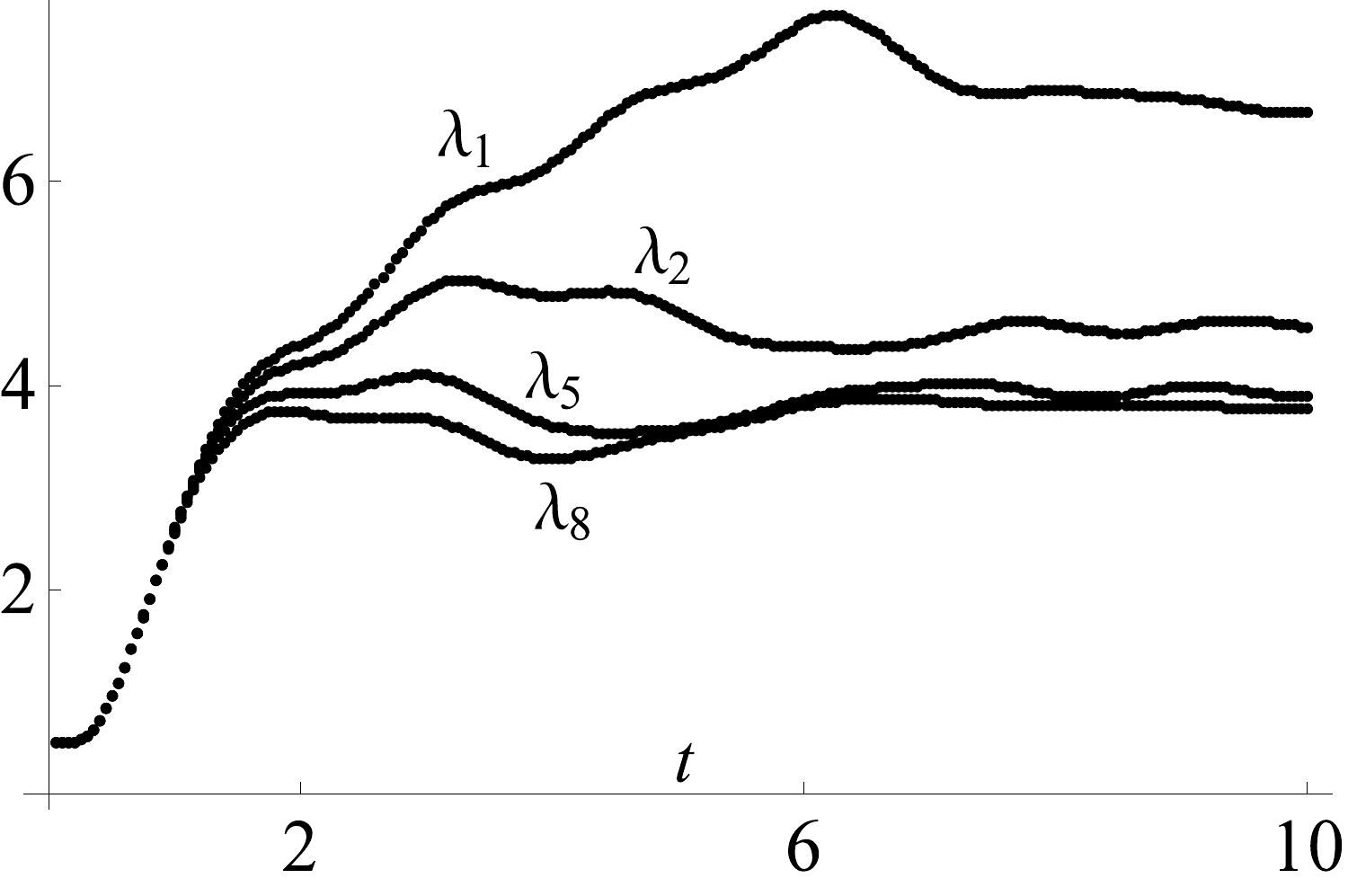}
\caption{(Color online.) Time evolution of the entanglement spectrum for the interacting quench with $L=10$. Left: Ordered entanglement eigenvalues at several times. An eigenvalue of $1/2$ corresponds to a completely unentangled mode. The jump at $n = 488 = 10^3 -8^3$ is a remanent of the ultra short time behavior. Only every twentieth eigenvalue is shown to improve readability. Right: The time evolution of the lowest eigenvalues. The eigenvalues $2,3,4$ and $5,6,7$ are each threefold degenerate. \label{fig:eigtime}}
\end{figure}

Data for the $\lambda_n$ for an $L=10$ system is shown in Fig.~\ref{fig:eigtime}, left panel. At $t=0$ (not shown), all correlations are completely localized and so there is no entanglement and all eigenvalues are $1/2$. As $t$ becomes greater than $L/2$, the light cone leaves the subsystem and the ES converges to a long time limit.  Similarly, the largest eigenvalue $\lambda_1$ (Fig.~\ref{fig:eigtime}, right panel) increases until $t\sim L/2$, while the smaller eigenvalues saturate at shorter times, as they correspond to shorter wavelengths. Superimposed on this underlying trend is an oscillation on the scale $t\sim 1$, presumably a reflection of the underlying lattice.

We now consider the entanglement entropy per unit volume $S(t)/L^3$ as a function of time, (Fig. \ref{fig:eigscaling}, bottom panels and Fig.~\ref{fig:entCombined2}). The basic shape of the curve reflects the light cone physics, increasing for $t < L/2$ and saturating $t \gg L/2$. We note three features: (i) the function
\begin{equation}
q(x) \equiv S(x L)/L^3, \quad x = t/(L)
\end{equation}
 converges to a well defined function as $L\rightarrow\infty$; (ii) up to a linear shift $q(x)$ depends only on the dispersion, and \emph{not} on whether the system  is free or interacting or on the energy; and (iii) the function $q(x)$ does not show the strict linear increase and kink at $L/2$ seen in $1d$ conformal systems~\cite{calabrese2007ent}.

\subsection{Quasi-Particle Picture}

The basic features of the entanglement entropy are explained by the quasi-particle picture. The quasi-particle approximation to the entanglement entropy is constructed as follows.
First at $t=0$ we imagine that all points in the system emit quasi-particles in all directions and at all momenta.
A pair of quasi-particles emitted from the same point at opposite momenta are assumed to be entangled and all others unentangled.
As the system is (quasi-)free these particles then fly at constant velocity.
At any given time $t$ we estimate the entanglement entropy by counting the number of entangled pairs where one quasi-particle is in the entanglement region and one quasi-particle is outside. Note that this argument is not able to fix an overall multiplicative constant since it does not calculate the entropy per entangled pair.

 The quasi-particle approximation immediately implies the the existence of a scaling function $q(x)$. This follows from the fact that the ballistic equation of motion $r = v_k t$, $v_k \equiv \partial_k \epsilon_k$  is invariant under simultaneous re-scaling of $r\rightarrow b r$ and $t\rightarrow b t$, for arbitrary $b$.  After such a rescaling all trajectories remain unchanged, however the volume of the entanglement region is multiplied by $b^d$. Therefore the number of entangled pairs in the region is multiplied by $b^d$ and since the entropy is supposed to be proportional to the number of pairs we obtain the invariance of $q(x)$.

 The function $q(x)$ depends only on the dispersion as the only input taken by the quasi-particle approximation is $v_k$. Therefore the leading part of the entanglement entropy is not sensitive to the universal exponents which define the long range physics.

  We do not expect a strict linear dependence as (i) we do not have a single velocity and (ii) even in the case of a single velocity, in the higher dimensional geometry we do not have a constant number of pairs entering the system.
\begin{figure}
\includegraphics[width =.49\columnwidth]{SIntUsualNew}
\includegraphics[width =.49\columnwidth]{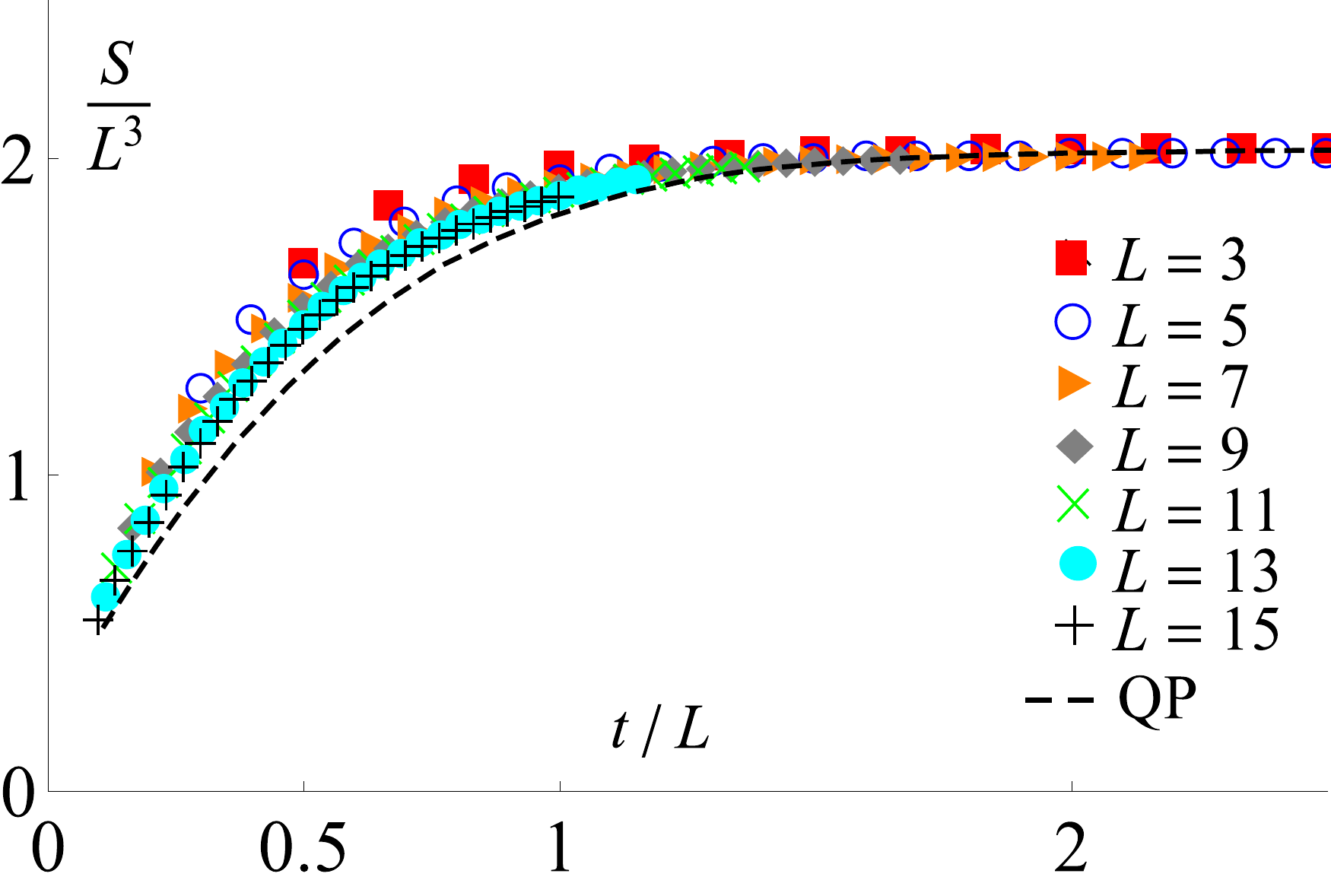}\\
\includegraphics[width =.49\columnwidth]{SIntAltNew}
\includegraphics[width =.49\columnwidth]{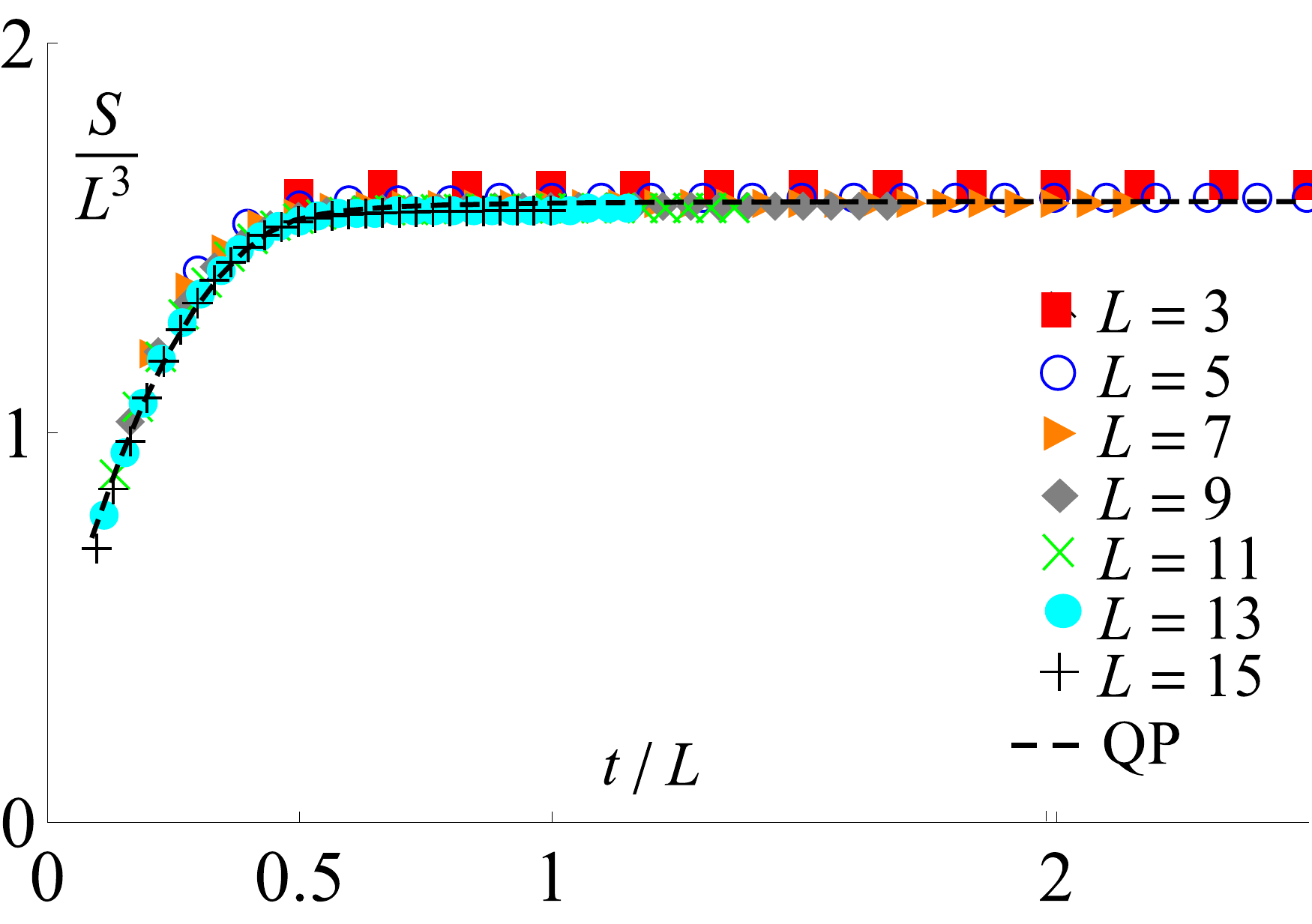}
\caption{Scaling plot of the entanglement entropy for the interacting quench n. n. dispersion (top left), free quench n. n. dispersion (top right), interacting quench, altered dispersion (bottom left) and free quench, altered dispersion (bottom right).
\label{fig:entCombined2} }
\end{figure}
To directly test this perspective, we calculate $q(x)$ by numerically simulating the motion of quasi-particles for the different dispersions. We calculate this by a  Monte Carlo algorithm. First we select a point $r$ at random from the interior of our region (in our case a cube, scaled to side length 1), and a momentum $k$. Then we calculate the velocity $v_k$ from the dispersion relation.  Next, we ask if the point $r + 2v_k t$ is inside the region. If it is not, that means there was a pair emitted at $r+ v_k t$, where one particle is in the region and one is outside, and thus contributing to the entropy. The proportion of tries where $r +2 v_k t$ is outside the region thus produces some function $F(x)$, which is zero for $t=0$ and goes to one as $t \rightarrow \infty$.

\begin{figure}
\includegraphics[width = \columnwidth]{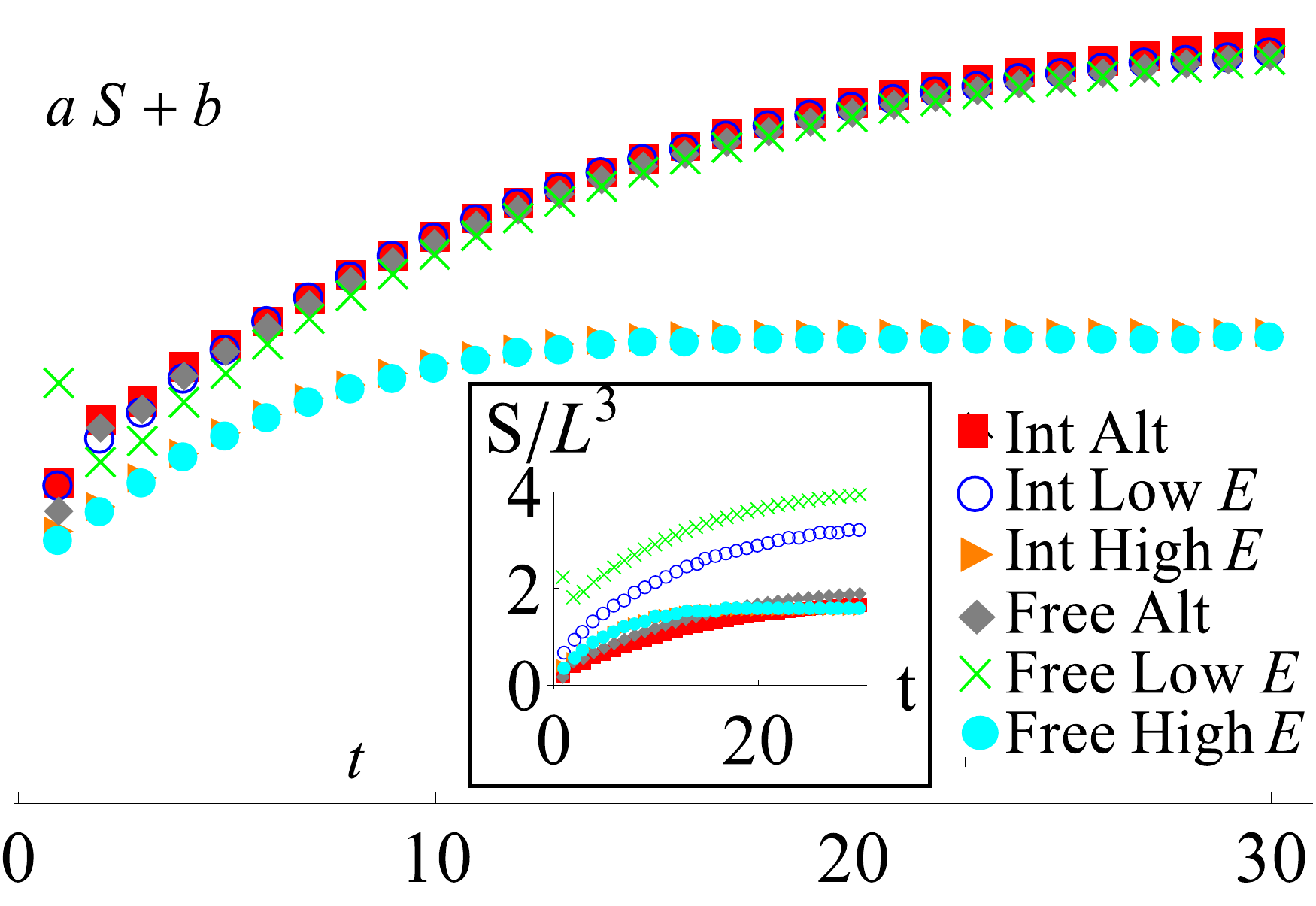}
\caption{(Color online.) Plot $a S(t) + b$ for six different quenches, with L = 15: low energy, high energy and alternate dispersion each for both the free and interacting case. Different fitting parameters $a$ and $b$. Note the curves collapse onto each other for quenches with the same dispersion. Inset: Unscaled entanglement entropy. \label{fig:entCombined} }
\end{figure}

 To compare this with the data at finite $L$, we use a fitting of the form $S(t)/L^3 \equiv a F\left(t/L\right) +b$. The fitting parameter $a$ measures the entropy per entangled particle, which is not fixed by the quasi-particle approximation. The parameter $b$ measures the amount of entropy generated at short times when $t$ is on the lattice scale, where we do not expect the continuum approximations made in the quasi-particle approximation to be valid. These fits are shown in Fig.~\ref{fig:entCombined2}. We find that up to a linear shift the approximation fits reasonably well, with some discrepancy that may be due to finite size corrections. The fact the entanglement entropy depends only on the dispersion, up to a linear shift, is shown in Fig.~\ref{fig:entCombined}.

\subsection{Scaling eigenvalues}

\begin{figure}
\includegraphics[width = .49\columnwidth]{FreeLambda1New}
\includegraphics[width = .49\columnwidth]{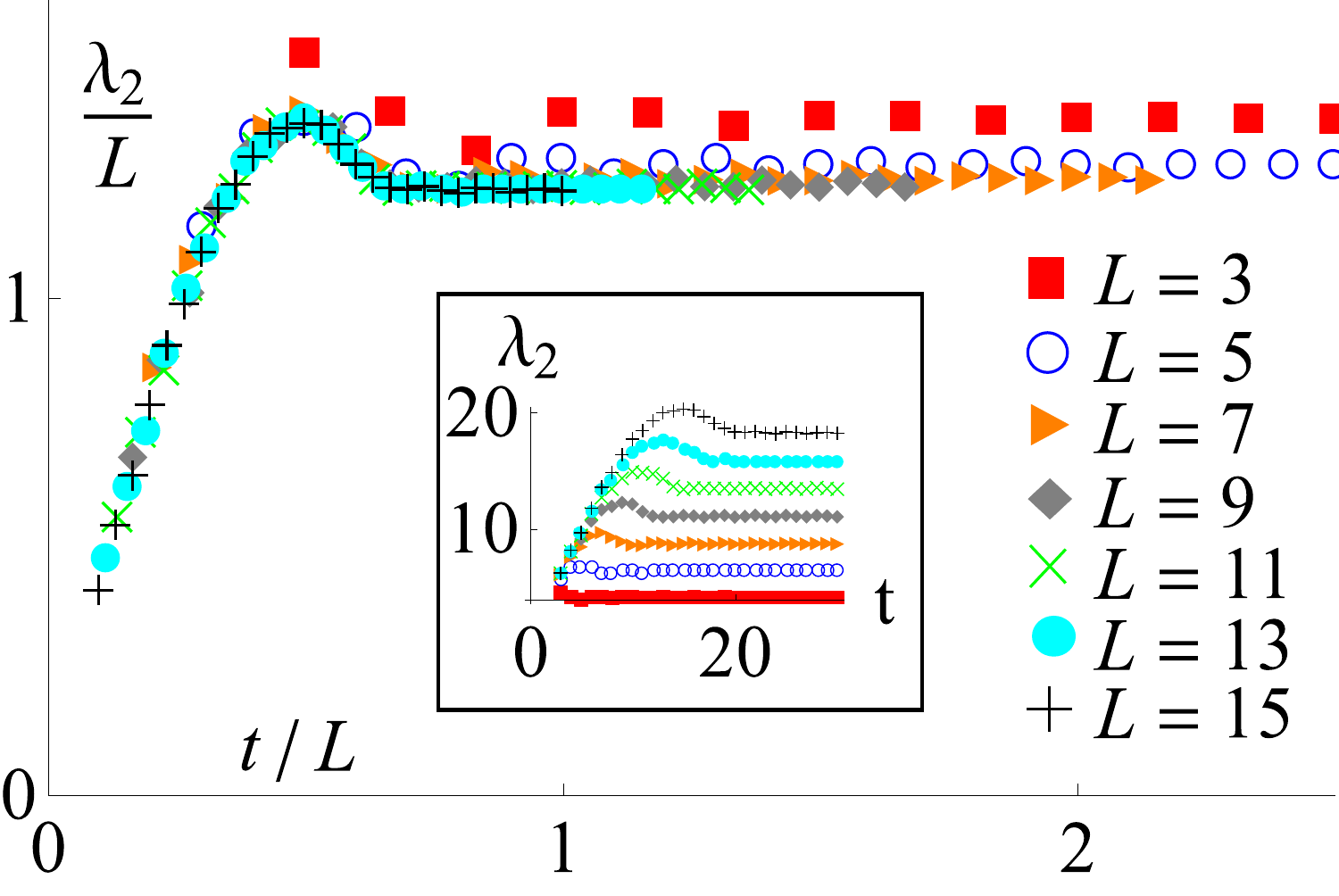}\\
\includegraphics[width = .49\columnwidth]{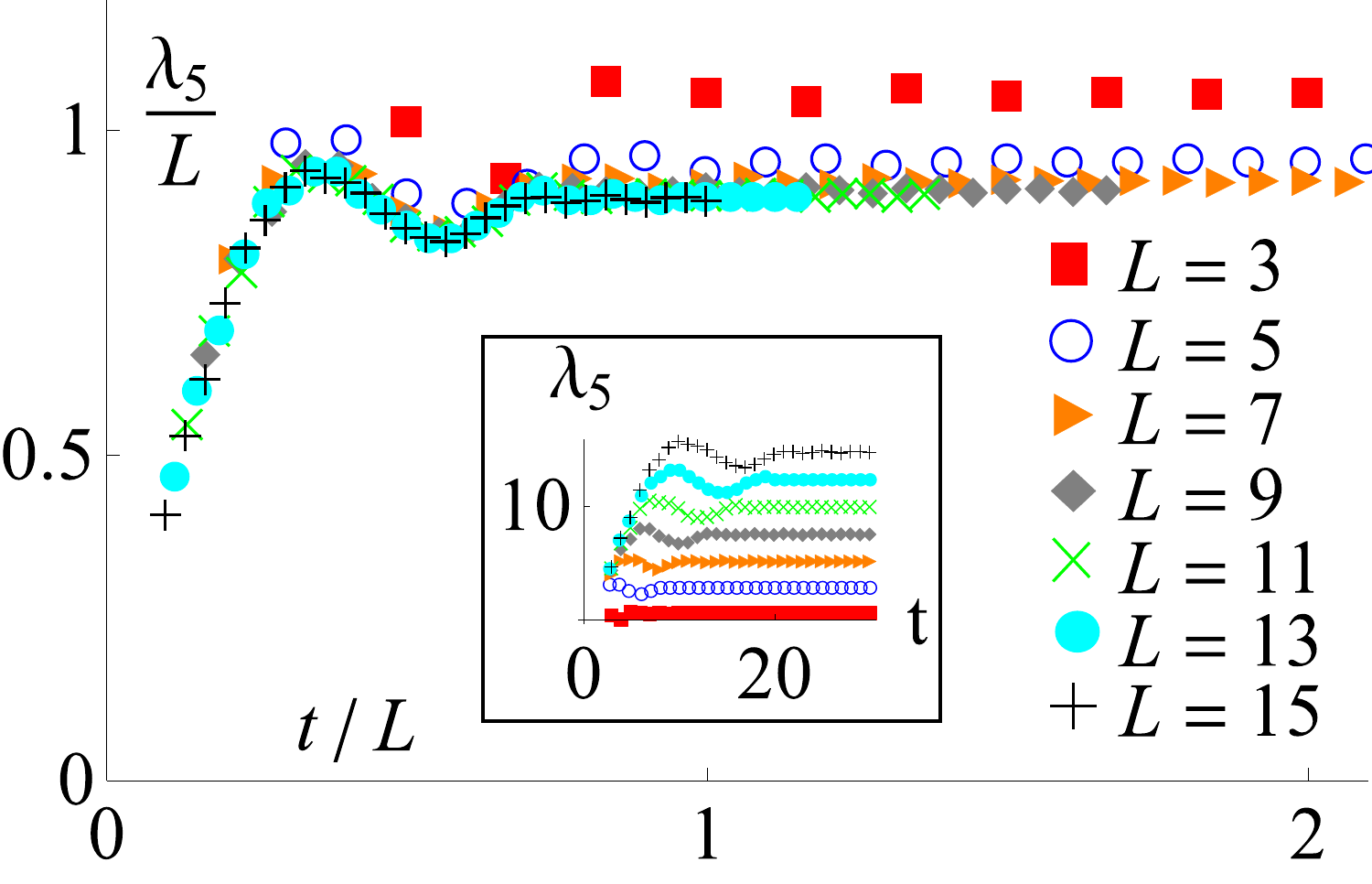}
\includegraphics[width = .49\columnwidth]{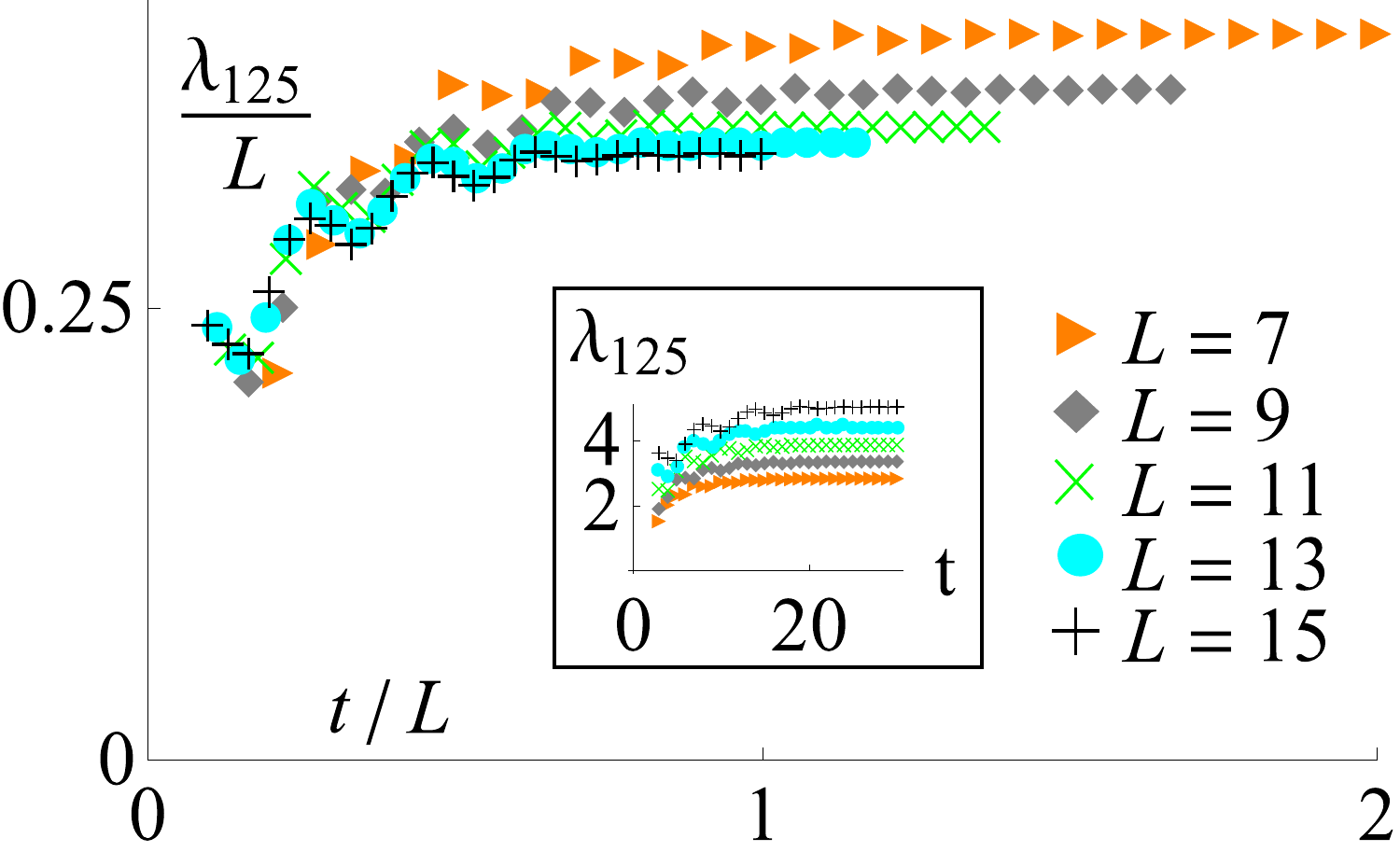}
\caption{(Color online.) Scaling plot of the $\lambda_1$, $\lambda_2$, $\lambda_5$ and $\lambda_{125}$ entanglement eigenvalues for the free, low energy quench, for various system lengths, over a time interval $t=0.5$ to $30$.  Inset: Unscaled data. The plot of $\lambda_{125}$ includes only larger system sizes and shows a worse fit.\label{fig:freeeigs}}
\end{figure}

\begin{figure}
\includegraphics[width = .49\columnwidth]{IntLambda1New}
\includegraphics[width = .49\columnwidth]{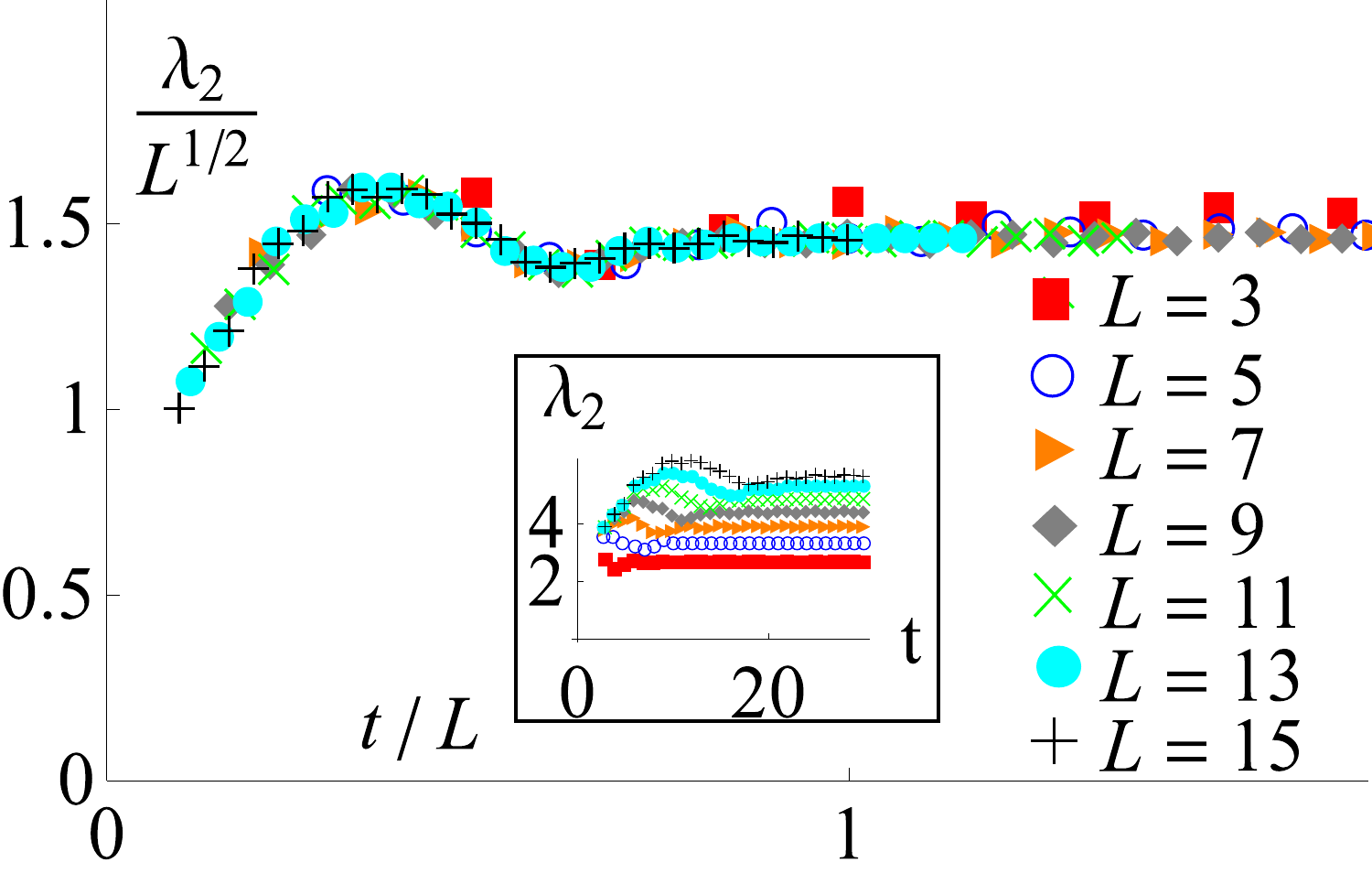}\\
\includegraphics[width = .49\columnwidth]{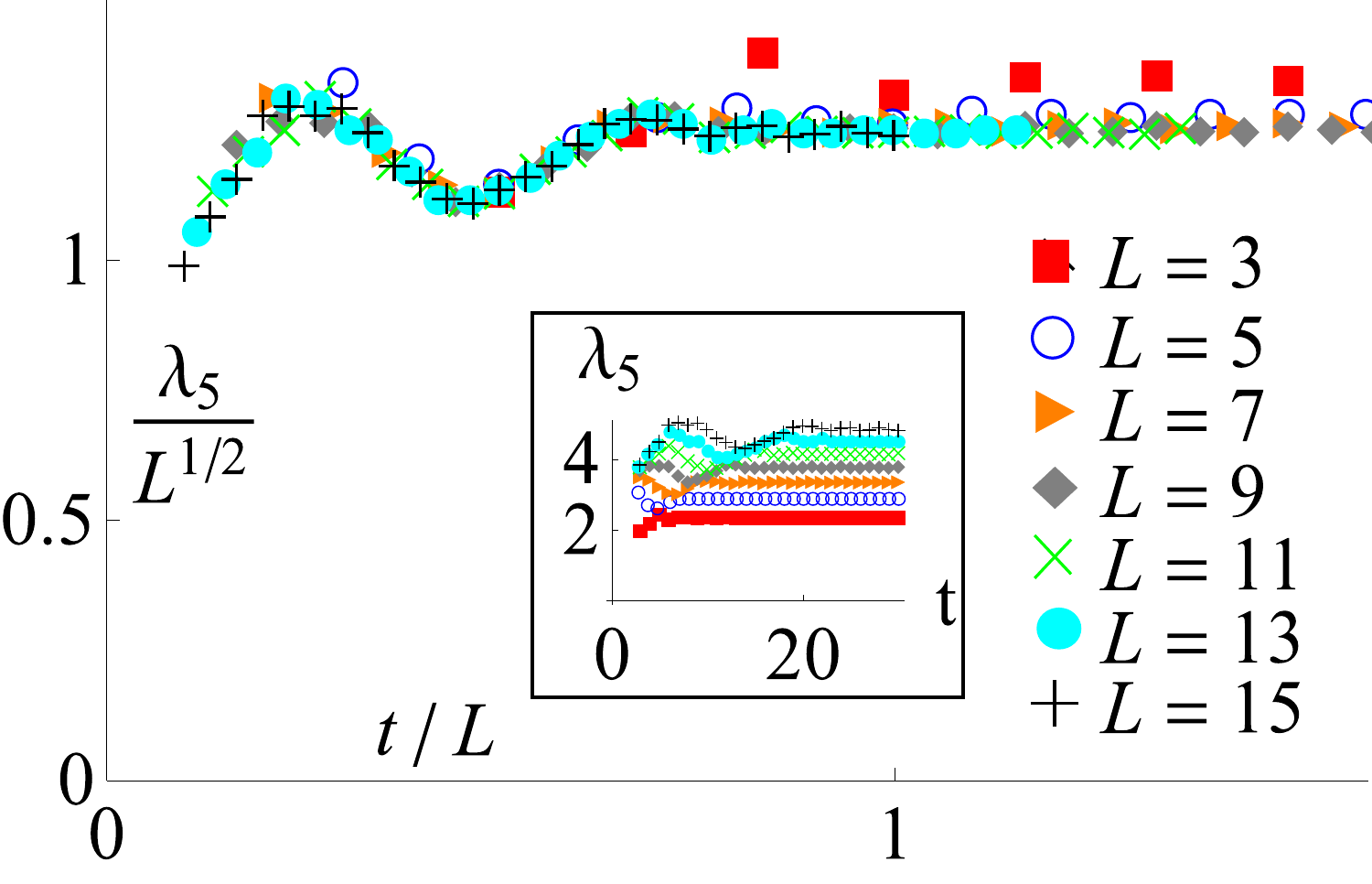}
\includegraphics[width = .49\columnwidth]{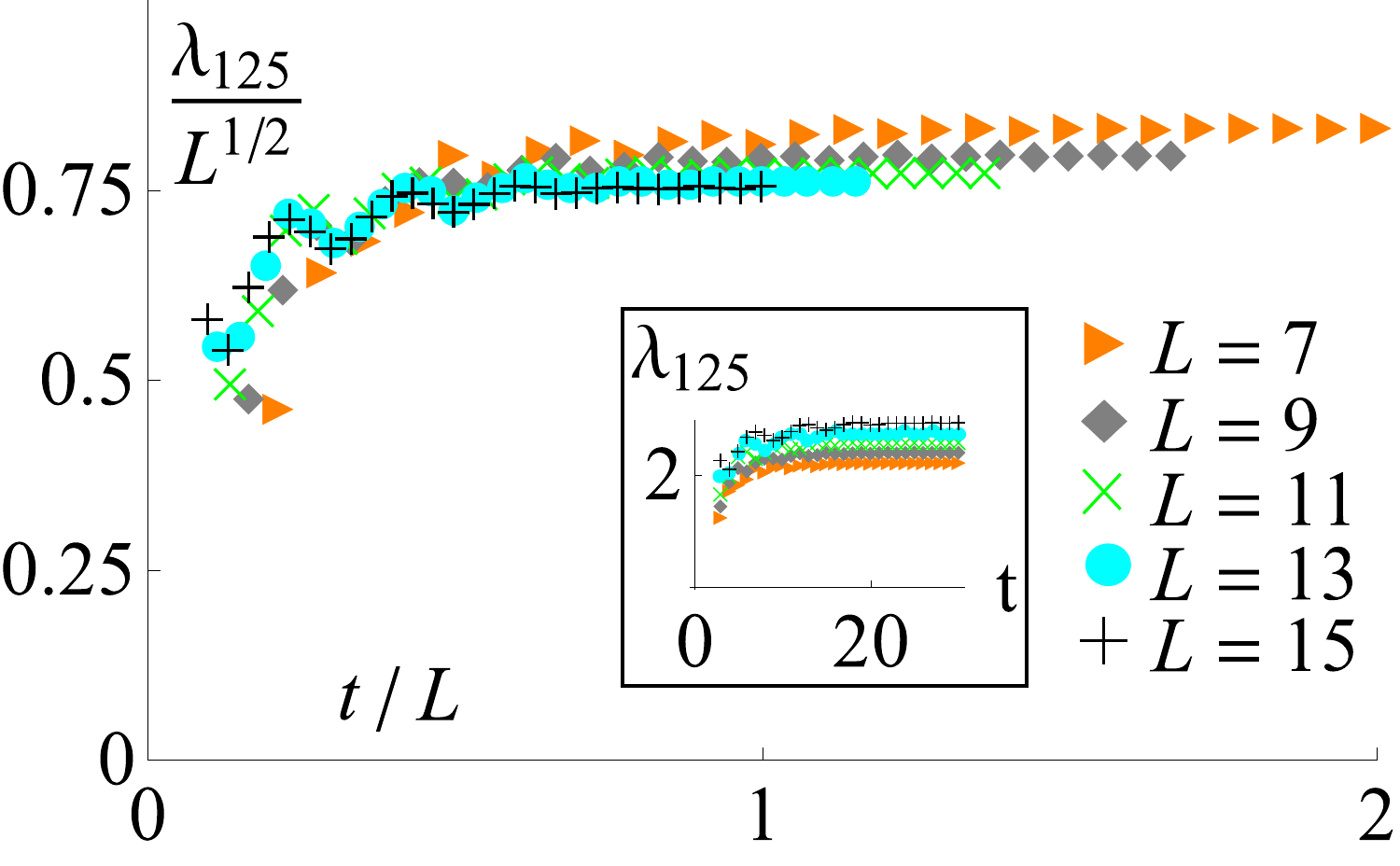}
\caption{(Color online.) Scaling plot of the $\lambda_1$, $\lambda_2$, $\lambda_5$ and $\lambda_{125}$ entanglement eigenvalues for the interacting, low energy quench, for various system lengths, over a time interval $t=.5$ to $30$.  Inset: Unscaled data. The plot of $\lambda_{125}$ includes only larger system sizes and shows a worse fit\label{fig:inteigs}}
\end{figure}
\begin{figure}
\includegraphics[width = .49\columnwidth]{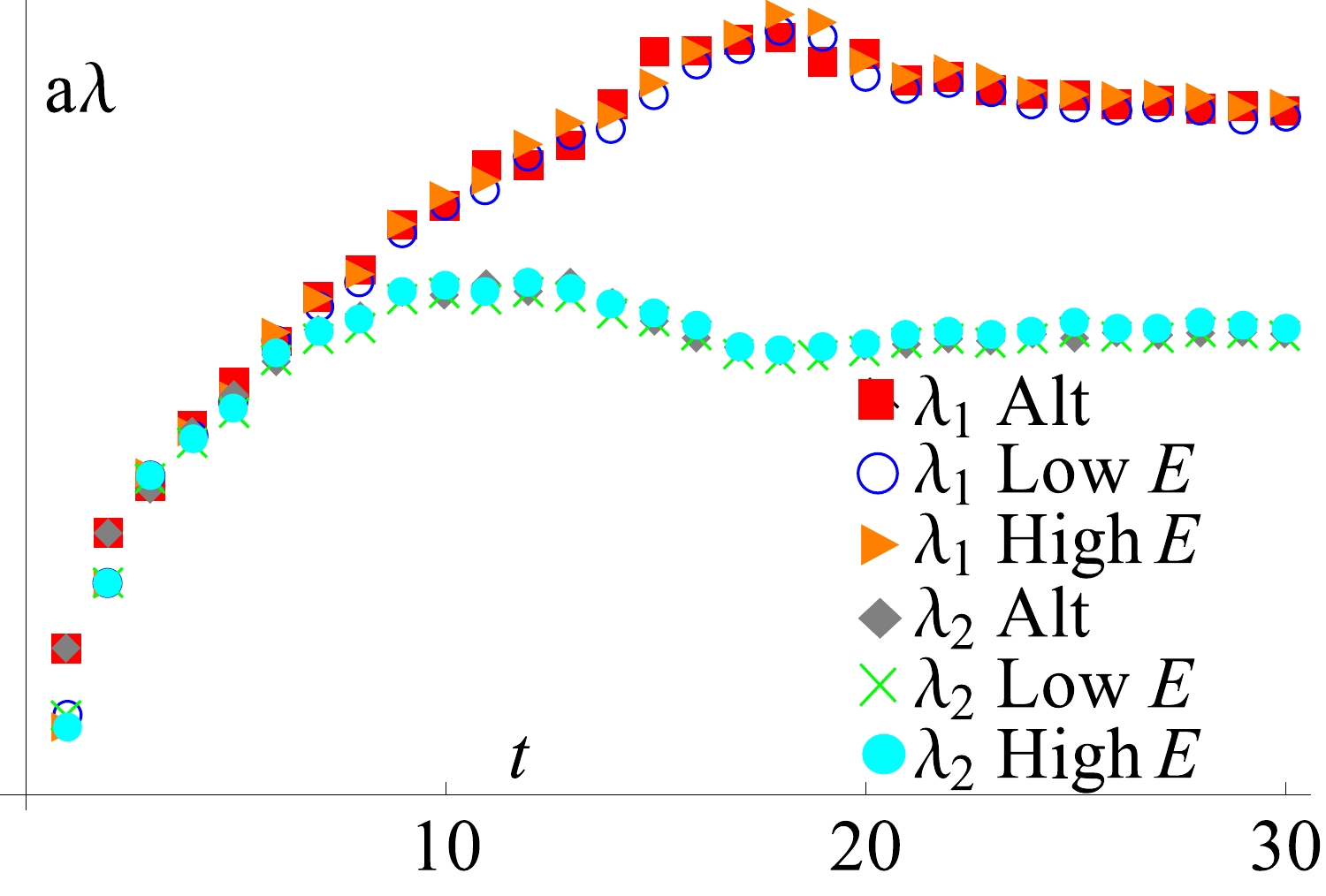}
\includegraphics[width = .49\columnwidth]{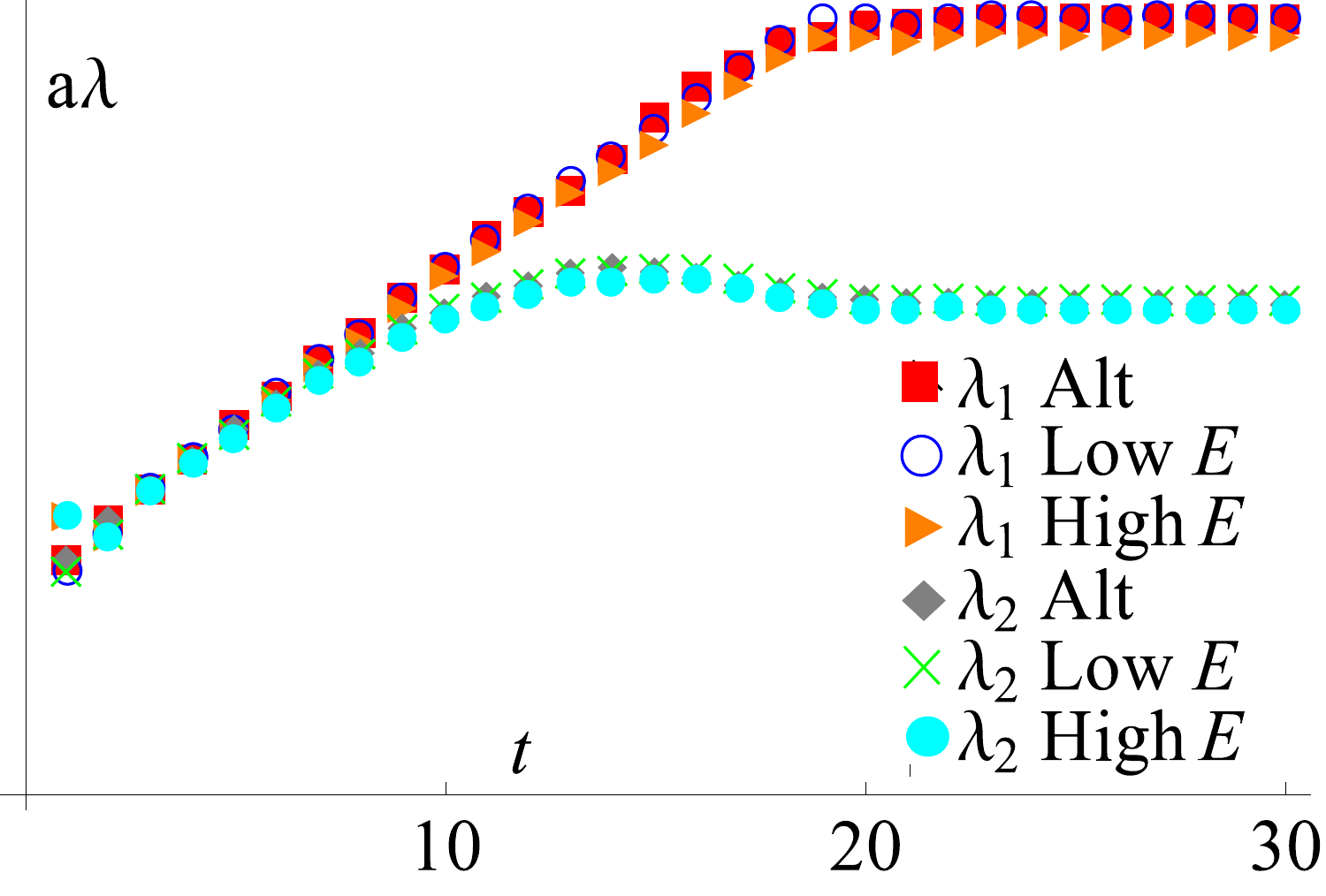}\\
\includegraphics[width = .49\columnwidth]{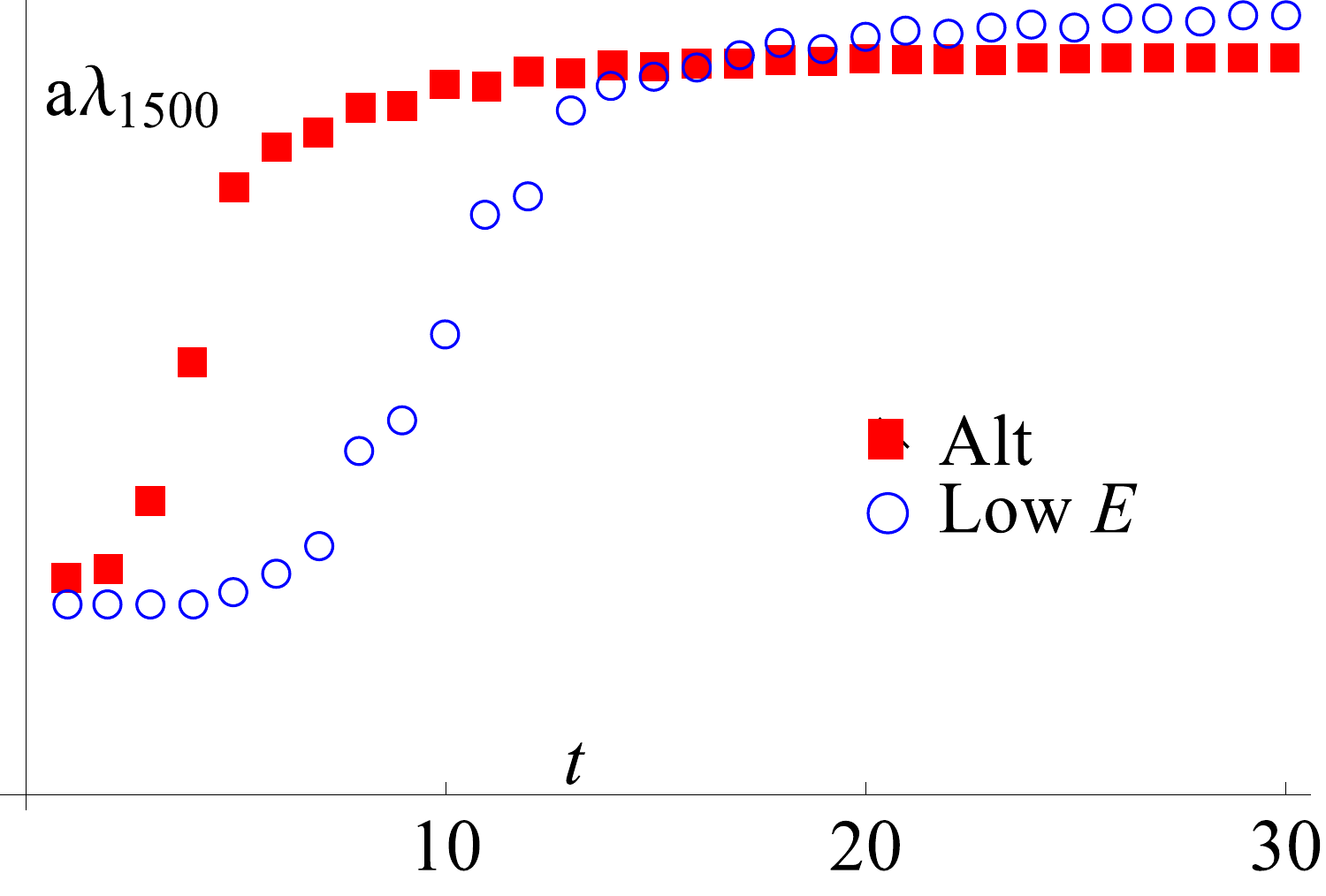}
\includegraphics[width = .49\columnwidth]{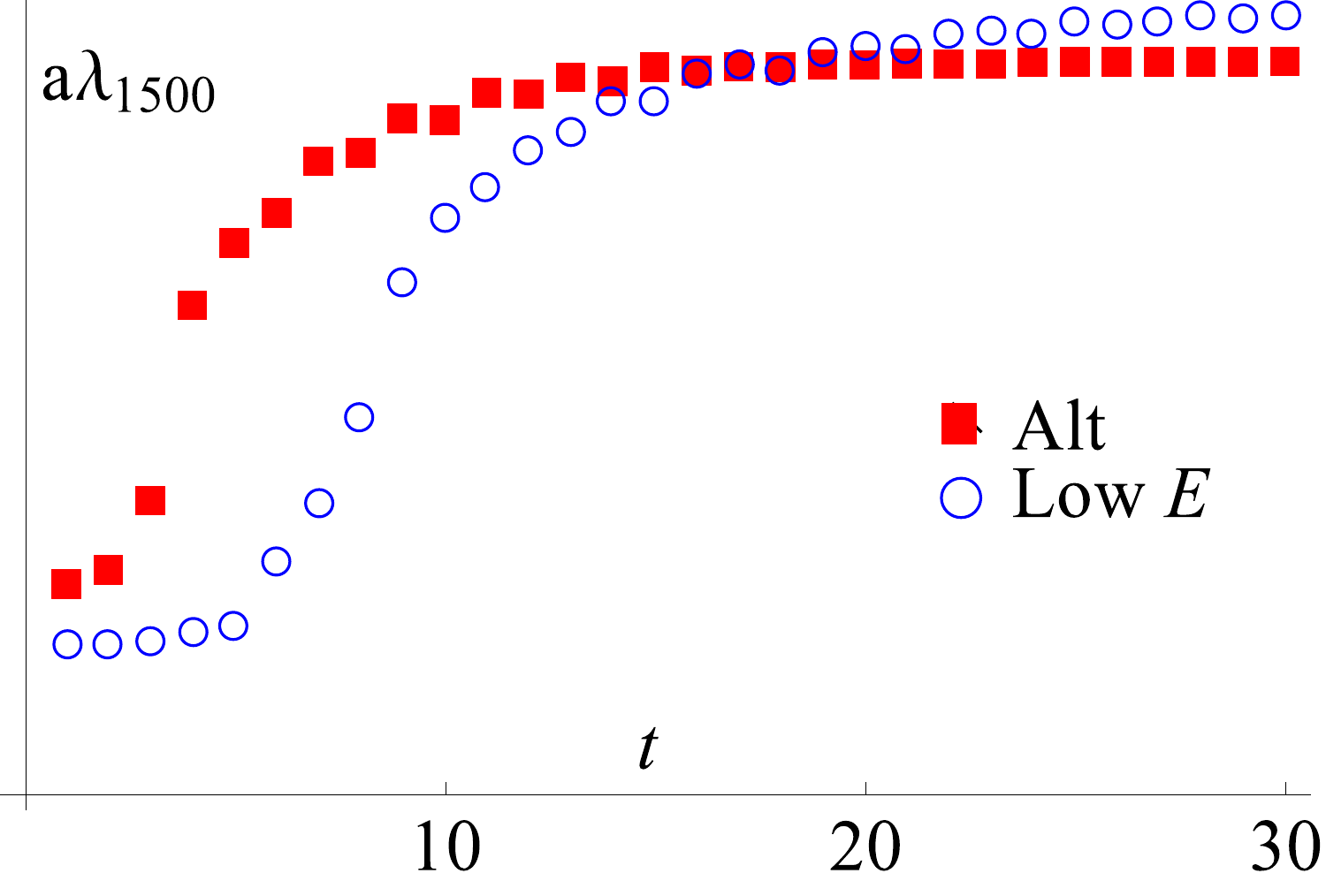}
\caption{(Color online.) Top: Comparison of the largest two eigenvalues for several different interacting critical (left) and free (right) quenches: the low energy, high energy, and alternate dispersion. The eigenvalues from each quench are multiplied by a fitting constant. Bottom : Comparison of the small eigenvalues for several different interacting critical (left) and free (right) quenches. Scaling the eigenvalues does not fit the curves.\label{fig:nonUniv}}
\end{figure}

We now argue that the largest eigenvalues scale according to the different critical exponents and are not sensitive to the short distance physics. These eigenvalues are given by the diagonalization of $\tilde{C}_{ij}(t)$, \niceref{eq:defCtilde}. We now make two assumptions about this diagonalization at large $L$: (i) the eigenvectors are smooth on the lattice scale so that a continuum limit may be taken and (ii) $C_{\mathcal{OO}'}$ may be replaced by their scaling forms. We emphasize that these are non-trivial assumptions, especially (ii) as it violates the commutation relation which is necessary for the constraint $\lambda > 1/2$. If these assumptions are valid, then the eigenvalue equation may be written as,
\begin{equation}
t^{2\alpha -d} \int_{L^d} d^d r'\hat{g}\left(\frac{r-r'}{t}\right)\xi(r')  =  \lambda(L,t) \xi(r),
\end{equation}
where $\hat{g}(x)$ is the matrix given by substituting the scaling forms for $C_{\mathcal{OO}'}$ into \niceref{eq:defCtilde}. By simultaneously scaling $r = \bar{r}L $,$r' = \bar{r}' L$ and $t = \bar{t} L$ we obtain the scaling form
\begin{equation}
\lambda(L,t) = L^{2\alpha} W(t/L).
\end{equation}

The largest eigenvalue for several different system sizes is plotted according to this scaling in Fig.~\ref{fig:eigscaling} for both the free ($\alpha = 1/2$, top left panel) and interacting ($\alpha = 1/4$ top right panel) cases. Interestingly a sharp feature is seen in this scaling curve for both systems at approximately $x \sim 0.6$.
Fig.~\ref{fig:freeeigs} and \ref{fig:inteigs} show that the scaling is effective for other large eigenvalues. Further we find that up to a linear re-scaling, the resulting curves are independent of the non-universal details of the quench for the largest eigenvalues, see Fig~\ref{fig:nonUniv}, top panels. However, both the scaling and the universality break down for smaller eigenvalues, see Fig~\ref{fig:nonUniv}, bottom panels.

 \subsection{Short-time behavior}
\begin{figure}
	\begin{center}	
	\includegraphics[width = .49\columnwidth]{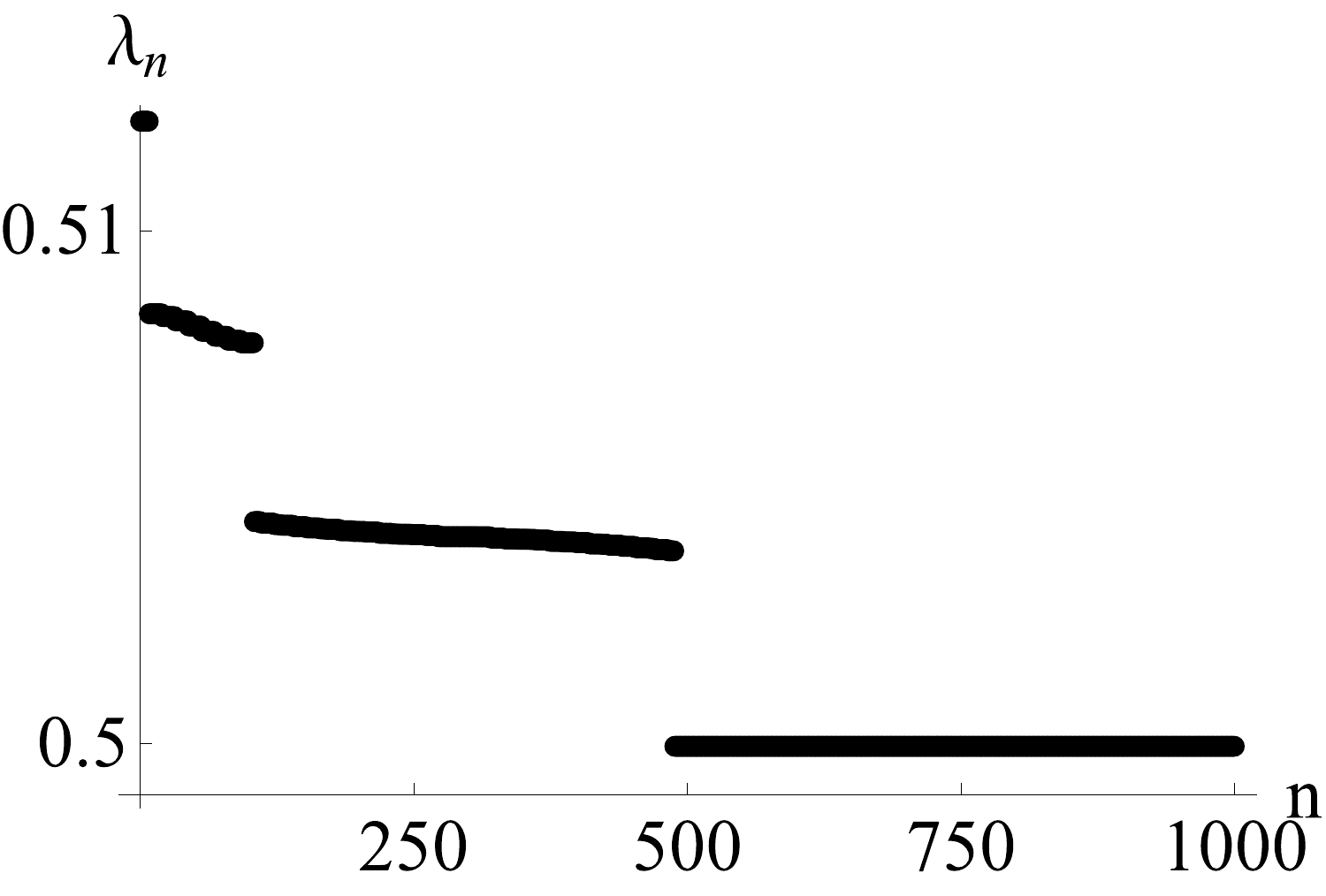}
	\end{center}
	\caption{
		The very short time structure of the entanglement spectrum for a free quench.
	\label{fig:ShortTime}
	}
\end{figure}

At very small times the entanglement spectra are dominated by short-range lattice effects. Results for $t <<1$ for an entanglement region of $L = 10$ are shown in Fig.~\ref{fig:ShortTime}.  Four equally spaced steps in the spectrum can be seen. Counting the number of points in each step gives 8, 96, 384 and 512. This suggests the points can be identified as corresponding to different positions of lattice points in the geometry: 8 corner points, 96 edge points, 384 face points, and 512 points in the interior. The eigenvalues are shifted from $1/2$ by an amount proportional to the number of of nearest neighbors in the complement: zero for the bulk points, one for the face points, two for the edge points and three for the corner points.

As $t$ increases the edge modes  rapidly hybridize into a continuous set of eigenvalues. However, the discontinuity between the edge states and the bulk states is visible until fairly late times, see Fig.~\ref{fig:eigtime}.

This behavior may be explained as follows. Let us consider the correlation matrix $\tilde{C}(t)$. At $t=0$ all correlations are non-zero except on the same lattice site. So $\tilde{C}$ is completely diagonal in position space. Now at $t>0$, this is no longer true. We may describe the evolution of $\tilde{C}$ by noting that the operators $X$ obey linear equations of motion. Therefore we may give their time evolution by 
\begin{equation}
X(t) = R(t)^T X(0),
\end{equation}
where $R(t)$  must be a symplectic matrix to conserve the commutation relations. Therefore the correlation matrix $\tilde{C}$, defined in Eq.~(\ref{eq:defCtilde}), evolves as
\begin{equation}
\tilde{C}(t) = \Omega R(t)^T {\bf C}(0) R(t) = R(t)^{-1}\tilde{C}(0)R(t).
\end{equation}

Now we wish to take the eigenvalues of $\tilde{C}$ when it is restricted to the entanglement region $A$.  Define the projector $P_A$  as giving $1$ on all vectors in Hilbert spaces in $A$ and $0$ on points outside and likewise $P_{\ba}\equiv 1- P_A$. 
 In terms of these therefore we want to diagonalize the operator
\begin{equation}
M \equiv  
	P_A \tilde{C}(t) P_A .
\end{equation}

Conjugating by $R(t)$, we see that this is equivalent to obtaining the eigenvalues of
\begin{align}
M'&\equiv  R(t) M R(t)^{-1} \nonumber\\
&= P_{A}(t) \tilde{C}(0) P_A(t),
\end{align}

where the time evolved projector is defined by
\begin{equation}
P_{A,\bar{A}}(t) = R(t)P_{A,\bar{A}}R(t)^{-1}
\end{equation}

Now $R(0)$ is the identity, and since the light cone spreads with speed $v = 1$, for $t \ll 1$, the matrix elements of $R(t)$ that connect different lattice sites are exponentially suppressed with increasing distance between the sites. 
Therefore it should be sufficient to retain only the matrix elements connecting the nearest neighbor sites. Call points in $A$ whose neighbors are all in $A$ bulk points, and otherwise boundary points. Because the part of $R$ that acts on bulk points of $R$ only connects points in $R$, it commutes with $P_{A,\bar{A}}$. Therefore the operator $P_{\bar{A}}(t)$ is zero on sites in the bulk of $A$ and one on sites in the bulk of  $\bar{A}$, and likewise for $P_A(t)$.

Since $\tilde{C}(0)$ is diagonal in position space, at short times the wavefunction which is supported on a single lattice site in the bulk of $A$, continues to be an eigenvector of 
$M'$ at short times with the same eigenvalue of $1/2$ as at $t=0$. 
For the points on the boundary this is not the case and the wavefunctions which are supported on the boundary will have their degeneracy lifted.

At the shortest times, we may Taylor expand
\begin{align}
 R(t) &\approx 1 - u t + \left (-v + \frac{1}{2}u^2\right)t^2 + \mathcal{O}(t^3)\\
 R(t)^{-1} &\approx 1 + u t + \left (v + \frac{1}{2}u^2\right)t^2 + \mathcal{O}(t^3)
\end{align}
 where $v$ and $u$ are matrices. We separate these matrices into parts 
 \begin{equation}
 u = u_{AA} + u_{\ba A} + u_{A \ba} + u_{\ba \ba}
 \end{equation} 
 and likewise for $v$, where $u_{AA} = P_A u P_{A}$, etc... With this decomposition we can write,
\begin{align}
 M 
  &= \biggr[ 1 + u_{AA} t + \left (v_{AA} + \frac{1}{2}u_{AA}^2\right)t^2\biggr] P_A \tilde{C}(0)P_A \nonumber\\
  &\quad\,\times  \biggr[ 1 - u_{AA} t + \left (-v_{AA} + \frac{1}{2}u_{AA}^2\right)t^2\biggr]\nonumber\\
  &
  \quad+
  \biggr[
  	\frac{1}{2}
   	\{
   	u_{A\ba}u_{\ba A}
  		,
  		\tilde{C}(0) 
	\}  	
  	+ 
  	u_{A\ba} 
  	\tilde{C}(0) 
  	u_{\ba A}
  \biggr]t^2\nonumber\\
&= R_{AA}^{-1} P_A\tilde{C}(0) P_AR_{AA}\nonumber\\
  &\quad+
  \biggr[
  	\frac{1}{2}
   	\{
   	u_{A\ba}u_{\ba A}
  		,
  		\tilde{C}(0) 
	\}  	
  	+ 
  	u_{A\ba} 
  	\tilde{C}(0) 
  	u_{\ba A}
  \biggr] t^2.
\end{align}
Here $R_{AA}$ and $R_{AA}^{-1}$ are defined by the terms in brackets in the previous equation.  It can be checked that to order $\mathcal{O}(t^3)$ they are in fact inverses of each other. Therefore we may conjugate $M$ by $R_{AA}$ without changing the eigenvalues and produce the matrix
\begin{align}
N
&\equiv R_{AA}M R_{AA}^{-1}\nonumber\\  
&= P_A \tilde{C}(0) P_A
 + R_{AA}
 \biggr[
  	\frac{1}{2}
   	\{
   	u_{A\ba}u_{\ba A}
  		,
  		\tilde{C}(0) 
	\}\nonumber\\  	
  	&\quad+ 
  	u_{A\ba} 
  	\tilde{C}(0) 
  	u_{\ba A}
 \biggr]
 R_{AA}^{-1}t^2
 \end{align}
 As we are working to order $t^2$ we may set $R_{AA}$ to $1$ in the proceeding equation. Therefore to the present order of approximation,
\begin{align}
N&\approx P_A \tilde{C}(0) P_A
 + 
 P_A
 \biggr[
  	\frac{1}{2}
   	\{
   	u_{A\ba}u_{\ba A}
  		,
  		\tilde{C}(0) 
	\}\nonumber\\  	
  	&\quad+ 
  	u_{A\ba} 
  	\tilde{C}(0) 
  	u_{\ba A}
 \biggr]
 P_A  
 t^2
\end{align}

 The first term $P_A \tilde{C}(0) P_A$ is simply the behavior at $t=0$ and is diagonal in position space with eigenvalues $1/2$. We now show that the second term is also diagonal in position space as long as we include only nearest neighbor hopping. The term $u_{A\ba}$ only connects nearest neighbor sites where one is in $A$ and the other in $\bar{A}$. 
Therefore it is only nonzero on the boundary points. In the entanglement geometry considered, every point in $\ba$ has at most one neighbor in $A$. 
Therefore the term $u_{A \ba}u_{\ba A}$ must connect a site on the boundary of $A$ with itself and therefore it is diagonal in position space. 
 
	Thus the eigenvectors of $N$ continue to be localized on a single lattice site. However, the eigenvalues are perturbatively corrected at order $t^2$ for each nearest neighbor $A$. As there is one correction for each nearest neighbor in $\ba$ and these are identical by the lattice symmetry, the splitting is strictly proportional to the number of nearest neighbors in $\ba$. This is the behavior seen in Fig.~\ref{fig:ShortTime}.

\section{Conclusion\label{sec:conclusion}} The entanglement spectrum was analyzed for quenches into both a free and critical interacting Hamiltonian.
 The largest eigenvalues were found to obey scaling laws determined by the critical exponents of the  system, while the entanglement entropy was found to be determined only by the details of the dispersion. The detection of the critical exponent $\theta$ means our results provide a connection between aging, important in localization and glasses, and entanglement.

 The results may be thought of as partitioning the ES into two parts.
 The largest eigenvalues correspond to long distance physics and carry information about the critical exponents of the system.
 The bulk of the entanglement spectrum carries information about the short-range physics and the non-universal details of the system.
 As the entanglement entropy averages over all eigenvalues it is largely determined by the bulk of the eigenvalues and therefore by the non-universal physics.

 We may interpret the density matrix $\rho$ as a thermal state $e^{-H_{\text{eff}}}$, with an effective Hamiltonian $H_{\text{eff}}$. A large eigenvalue then corresponds to a highly occupied and therefore low energy mode of $H_{\text{eff}}$.
The partition of the eigenvalues means the lowest energy modes of  $H_{\text{eff}}$ correspond to long-range universal physics, just as for the physical Hamiltonian.

The study of the ES
as an alternate understanding of correlated quantum ground states is an active new direction of research~\cite{Haldane08}. Our work provides guidance for studying this in non-equilibrium by highlighting the importance of criticality and the structure of the ES.

The $N\rightarrow\infty$ limit could be improved by considering $1/N$ corrections~\cite{Berges2003}. Although this would provide only perturbative corrections to the critical regime studied, it might allow for studying the ES during the crossover between the critical and non-universal thermal phases. However, as no known method for a $1/N$ expansion of the ES exists, this would be a significant challenge.

We note that the scaling of the first hundred eigenvalues is well reproduced by the numerical results even for $L <10$. At these lengths it would be non-trivial to extract the critical exponents from the correlation functions, Fig.~\ref{fig:Correlators}. Moreover, the scaling can be tested independently on a large number of eigenvalues. Therefore it is worth investigating whether the ES is an efficient means to extract critical exponents in systems where the critical exponents are not known.

\begin{acknowledgments}
The authors are grateful to A. Chiocchetta and A. Gambassi for helpful discussions.
This work was supported by US National Science Foundation Grant NSF-DMR 1303177.
\end{acknowledgments}


\begin{thebibliography}{38}%
\makeatletter
\providecommand \@ifxundefined [1]{%
 \@ifx{#1\undefined}
}%
\providecommand \@ifnum [1]{%
 \ifnum #1\expandafter \@firstoftwo
 \else \expandafter \@secondoftwo
 \fi
}%
\providecommand \@ifx [1]{%
 \ifx #1\expandafter \@firstoftwo
 \else \expandafter \@secondoftwo
 \fi
}%
\providecommand \natexlab [1]{#1}%
\providecommand \enquote  [1]{``#1''}%
\providecommand \bibnamefont  [1]{#1}%
\providecommand \bibfnamefont [1]{#1}%
\providecommand \citenamefont [1]{#1}%
\providecommand \href@noop [0]{\@secondoftwo}%
\providecommand \href [0]{\begingroup \@sanitize@url \@href}%
\providecommand \@href[1]{\@@startlink{#1}\@@href}%
\providecommand \@@href[1]{\endgroup#1\@@endlink}%
\providecommand \@sanitize@url [0]{\catcode `\\12\catcode `\$12\catcode
  `\&12\catcode `\#12\catcode `\^12\catcode `\_12\catcode `\%12\relax}%
\providecommand \@@startlink[1]{}%
\providecommand \@@endlink[0]{}%
\providecommand \url  [0]{\begingroup\@sanitize@url \@url }%
\providecommand \@url [1]{\endgroup\@href {#1}{\urlprefix }}%
\providecommand \urlprefix  [0]{URL }%
\providecommand \Eprint [0]{\href }%
\providecommand \doibase [0]{http://dx.doi.org/}%
\providecommand \selectlanguage [0]{\@gobble}%
\providecommand \bibinfo  [0]{\@secondoftwo}%
\providecommand \bibfield  [0]{\@secondoftwo}%
\providecommand \translation [1]{[#1]}%
\providecommand \BibitemOpen [0]{}%
\providecommand \bibitemStop [0]{}%
\providecommand \bibitemNoStop [0]{.\EOS\space}%
\providecommand \EOS [0]{\spacefactor3000\relax}%
\providecommand \BibitemShut  [1]{\csname bibitem#1\endcsname}%
\let\auto@bib@innerbib\@empty
\bibitem [{\citenamefont {Calabrese}\ and\ \citenamefont
  {Cardy}(2006)}]{Calabrese2006}%
  \BibitemOpen
  \bibfield  {author} {\bibinfo {author} {\bibfnamefont {P.}~\bibnamefont
  {Calabrese}}\ and\ \bibinfo {author} {\bibfnamefont {J.}~\bibnamefont
  {Cardy}},\ }\href {\doibase 10.1103/PhysRevLett.96.136801} {\bibfield
  {journal} {\bibinfo  {journal} {Phys. Rev. Lett.}\ }\textbf {\bibinfo
  {volume} {96}},\ \bibinfo {pages} {136801} (\bibinfo {year}
  {2006})}\BibitemShut {NoStop}%
\bibitem [{\citenamefont {Calabrese}\ and\ \citenamefont
  {Cardy}(2007{\natexlab{a}})}]{Calabrese2007}%
  \BibitemOpen
  \bibfield  {author} {\bibinfo {author} {\bibfnamefont {P.}~\bibnamefont
  {Calabrese}}\ and\ \bibinfo {author} {\bibfnamefont {J.}~\bibnamefont
  {Cardy}},\ }\href {http://stacks.iop.org/1742-5468/2007/i=06/a=P06008}
  {\bibfield  {journal} {\bibinfo  {journal} {J. Stat. Mech.}\ }\textbf
  {\bibinfo {volume} {06}},\ \bibinfo {pages} {P06008} (\bibinfo {year}
  {2007}{\natexlab{a}})}\BibitemShut {NoStop}%
\bibitem [{\citenamefont {Polkovnikov}\ \emph {et~al.}(2011)\citenamefont
  {Polkovnikov}, \citenamefont {Sengupta}, \citenamefont {Silva},\ and\
  \citenamefont {Vengalattore}}]{PolkovnikovRMP}%
  \BibitemOpen
  \bibfield  {author} {\bibinfo {author} {\bibfnamefont {A.}~\bibnamefont
  {Polkovnikov}}, \bibinfo {author} {\bibfnamefont {K.}~\bibnamefont
  {Sengupta}}, \bibinfo {author} {\bibfnamefont {A.}~\bibnamefont {Silva}}, \
  and\ \bibinfo {author} {\bibfnamefont {M.}~\bibnamefont {Vengalattore}},\
  }\href {\doibase 10.1103/RevModPhys.83.863} {\bibfield  {journal} {\bibinfo
  {journal} {Rev. Mod. Phys.}\ }\textbf {\bibinfo {volume} {83}},\ \bibinfo
  {pages} {863} (\bibinfo {year} {2011})}\BibitemShut {NoStop}%
\bibitem [{\citenamefont {Bloch}\ \emph {et~al.}(2008)\citenamefont {Bloch},
  \citenamefont {Dalibard},\ and\ \citenamefont {Zwerger}}]{Bloch2008}%
  \BibitemOpen
  \bibfield  {author} {\bibinfo {author} {\bibfnamefont {I.}~\bibnamefont
  {Bloch}}, \bibinfo {author} {\bibfnamefont {J.}~\bibnamefont {Dalibard}}, \
  and\ \bibinfo {author} {\bibfnamefont {W.}~\bibnamefont {Zwerger}},\ }\href
  {\doibase 10.1103/RevModPhys.80.885} {\bibfield  {journal} {\bibinfo
  {journal} {Rev. Mod. Phys.}\ }\textbf {\bibinfo {volume} {80}},\ \bibinfo
  {pages} {885} (\bibinfo {year} {2008})}\BibitemShut {NoStop}%
\bibitem [{\citenamefont {Trotzky}\ \emph {et~al.}(2012)\citenamefont
  {Trotzky}, \citenamefont {Chen}, \citenamefont {Flesch}, \citenamefont
  {McCulloch}, \citenamefont {Schollw\"ock}, \citenamefont {Eisert},\ and\
  \citenamefont {Bloch}}]{Trotzky2012}%
  \BibitemOpen
  \bibfield  {author} {\bibinfo {author} {\bibfnamefont {S.}~\bibnamefont
  {Trotzky}}, \bibinfo {author} {\bibfnamefont {Y.-A.}\ \bibnamefont {Chen}},
  \bibinfo {author} {\bibfnamefont {A.}~\bibnamefont {Flesch}}, \bibinfo
  {author} {\bibfnamefont {I.~P.}\ \bibnamefont {McCulloch}}, \bibinfo {author}
  {\bibfnamefont {U.}~\bibnamefont {Schollw\"ock}}, \bibinfo {author}
  {\bibfnamefont {J.}~\bibnamefont {Eisert}}, \ and\ \bibinfo {author}
  {\bibfnamefont {I.}~\bibnamefont {Bloch}},\ }\href@noop {} {\bibfield
  {journal} {\bibinfo  {journal} {Nature Physics}\ }\textbf {\bibinfo {volume}
  {8}},\ \bibinfo {pages} {325} (\bibinfo {year} {2012})}\BibitemShut {NoStop}%
\bibitem [{\citenamefont {Gring}\ \emph {et~al.}(2012)\citenamefont {Gring},
  \citenamefont {Kuhnert}, \citenamefont {Langen}, \citenamefont {Kitagawa},
  \citenamefont {Rauer}, \citenamefont {Schreitl}, \citenamefont {Mazets},
  \citenamefont {Smith}, \citenamefont {Demler},\ and\ \citenamefont
  {Schmiedmayer}}]{Gring2012}%
  \BibitemOpen
  \bibfield  {author} {\bibinfo {author} {\bibfnamefont {M.}~\bibnamefont
  {Gring}}, \bibinfo {author} {\bibfnamefont {M.}~\bibnamefont {Kuhnert}},
  \bibinfo {author} {\bibfnamefont {T.}~\bibnamefont {Langen}}, \bibinfo
  {author} {\bibfnamefont {T.}~\bibnamefont {Kitagawa}}, \bibinfo {author}
  {\bibfnamefont {B.}~\bibnamefont {Rauer}}, \bibinfo {author} {\bibfnamefont
  {M.}~\bibnamefont {Schreitl}}, \bibinfo {author} {\bibfnamefont
  {I.}~\bibnamefont {Mazets}}, \bibinfo {author} {\bibfnamefont {D.~A.}\
  \bibnamefont {Smith}}, \bibinfo {author} {\bibfnamefont {E.}~\bibnamefont
  {Demler}}, \ and\ \bibinfo {author} {\bibfnamefont {J.}~\bibnamefont
  {Schmiedmayer}},\ }\href@noop {} {\bibfield  {journal} {\bibinfo  {journal}
  {Science}\ }\textbf {\bibinfo {volume} {337}},\ \bibinfo {pages} {1318}
  (\bibinfo {year} {2012})}\BibitemShut {NoStop}%
\bibitem [{\citenamefont {Langen}\ \emph {et~al.}(2013)\citenamefont {Langen},
  \citenamefont {Geiger}, \citenamefont {Kuhnert}, \citenamefont {Rauer},\ and\
  \citenamefont {Schmiedmayer}}]{Langen2013}%
  \BibitemOpen
  \bibfield  {author} {\bibinfo {author} {\bibfnamefont {T.}~\bibnamefont
  {Langen}}, \bibinfo {author} {\bibfnamefont {R.}~\bibnamefont {Geiger}},
  \bibinfo {author} {\bibfnamefont {M.}~\bibnamefont {Kuhnert}}, \bibinfo
  {author} {\bibfnamefont {B.}~\bibnamefont {Rauer}}, \ and\ \bibinfo {author}
  {\bibfnamefont {J.}~\bibnamefont {Schmiedmayer}},\ }\href@noop {} {\bibfield
  {journal} {\bibinfo  {journal} {Nature Phys.}\ }\textbf {\bibinfo {volume}
  {9}},\ \bibinfo {pages} {640} (\bibinfo {year} {2013})}\BibitemShut {NoStop}%
\bibitem [{\citenamefont {Richerme}\ \emph {et~al.}(2014)\citenamefont
  {Richerme}, \citenamefont {Gong}, \citenamefont {Lee}, \citenamefont {Senko},
  \citenamefont {Smith}, \citenamefont {Foss-Feig}, \citenamefont {Michalakis},
  \citenamefont {Gorshkov},\ and\ \citenamefont {Monroe}}]{Monroe14}%
  \BibitemOpen
  \bibfield  {author} {\bibinfo {author} {\bibfnamefont {P.}~\bibnamefont
  {Richerme}}, \bibinfo {author} {\bibfnamefont {Z.-X.}\ \bibnamefont {Gong}},
  \bibinfo {author} {\bibfnamefont {A.}~\bibnamefont {Lee}}, \bibinfo {author}
  {\bibfnamefont {C.}~\bibnamefont {Senko}}, \bibinfo {author} {\bibfnamefont
  {J.}~\bibnamefont {Smith}}, \bibinfo {author} {\bibfnamefont
  {M.}~\bibnamefont {Foss-Feig}}, \bibinfo {author} {\bibfnamefont
  {S.}~\bibnamefont {Michalakis}}, \bibinfo {author} {\bibfnamefont {A.~V.}\
  \bibnamefont {Gorshkov}}, \ and\ \bibinfo {author} {\bibfnamefont
  {C.}~\bibnamefont {Monroe}},\ }\href@noop {} {\bibfield  {journal} {\bibinfo
  {journal} {Nature Letter}\ }\textbf {\bibinfo {volume} {511}},\ \bibinfo
  {pages} {198} (\bibinfo {year} {2014})}\BibitemShut {NoStop}%
\bibitem [{\citenamefont {Vidal}\ \emph {et~al.}(2003)\citenamefont {Vidal},
  \citenamefont {Latorre}, \citenamefont {Rico},\ and\ \citenamefont
  {Kitaev}}]{Vidal03}%
  \BibitemOpen
  \bibfield  {author} {\bibinfo {author} {\bibfnamefont {G.}~\bibnamefont
  {Vidal}}, \bibinfo {author} {\bibfnamefont {J.~I.}\ \bibnamefont {Latorre}},
  \bibinfo {author} {\bibfnamefont {E.}~\bibnamefont {Rico}}, \ and\ \bibinfo
  {author} {\bibfnamefont {A.}~\bibnamefont {Kitaev}},\ }\href {\doibase
  10.1103/PhysRevLett.90.227902} {\bibfield  {journal} {\bibinfo  {journal}
  {Phys. Rev. Lett.}\ }\textbf {\bibinfo {volume} {90}},\ \bibinfo {pages}
  {227902} (\bibinfo {year} {2003})}\BibitemShut {NoStop}%
\bibitem [{\citenamefont {Vidal}(2007)}]{Vidal07}%
  \BibitemOpen
  \bibfield  {author} {\bibinfo {author} {\bibfnamefont {G.}~\bibnamefont
  {Vidal}},\ }\href {\doibase 10.1103/PhysRevLett.99.220405} {\bibfield
  {journal} {\bibinfo  {journal} {Phys. Rev. Lett.}\ }\textbf {\bibinfo
  {volume} {99}},\ \bibinfo {pages} {220405} (\bibinfo {year}
  {2007})}\BibitemShut {NoStop}%
\bibitem [{\citenamefont {Casini}\ and\ \citenamefont
  {Huerta}(2007)}]{Casini07}%
  \BibitemOpen
  \bibfield  {author} {\bibinfo {author} {\bibfnamefont {H.}~\bibnamefont
  {Casini}}\ and\ \bibinfo {author} {\bibfnamefont {M.}~\bibnamefont
  {Huerta}},\ }\href {\doibase
  http://dx.doi.org/10.1016/j.nuclphysb.2006.12.012} {\bibfield  {journal}
  {\bibinfo  {journal} {Nuclear Physics B}\ }\textbf {\bibinfo {volume}
  {764}},\ \bibinfo {pages} {183 } (\bibinfo {year} {2007})}\BibitemShut
  {NoStop}%
\bibitem [{\citenamefont {Casini}\ and\ \citenamefont
  {Huerta}(2009)}]{Casini09}%
  \BibitemOpen
  \bibfield  {author} {\bibinfo {author} {\bibfnamefont {H.}~\bibnamefont
  {Casini}}\ and\ \bibinfo {author} {\bibfnamefont {M.}~\bibnamefont
  {Huerta}},\ }\href {http://stacks.iop.org/1751-8121/42/i=50/a=504007}
  {\bibfield  {journal} {\bibinfo  {journal} {Journal of Physics A:
  Mathematical and Theoretical}\ }\textbf {\bibinfo {volume} {42}},\ \bibinfo
  {pages} {504007} (\bibinfo {year} {2009})}\BibitemShut {NoStop}%
\bibitem [{\citenamefont {Eisert}\ \emph {et~al.}(2010)\citenamefont {Eisert},
  \citenamefont {Cramer},\ and\ \citenamefont {Plenio}}]{EntRmp10}%
  \BibitemOpen
  \bibfield  {author} {\bibinfo {author} {\bibfnamefont {J.}~\bibnamefont
  {Eisert}}, \bibinfo {author} {\bibfnamefont {M.}~\bibnamefont {Cramer}}, \
  and\ \bibinfo {author} {\bibfnamefont {M.~B.}\ \bibnamefont {Plenio}},\
  }\href {\doibase 10.1103/RevModPhys.82.277} {\bibfield  {journal} {\bibinfo
  {journal} {Rev. Mod. Phys.}\ }\textbf {\bibinfo {volume} {82}},\ \bibinfo
  {pages} {277} (\bibinfo {year} {2010})}\BibitemShut {NoStop}%
\bibitem [{\citenamefont {Zeng}\ \emph {et~al.}(2015)\citenamefont {Zeng},
  \citenamefont {Chen}, \citenamefont {Zhou},\ and\ \citenamefont
  {Wen}}]{Wenbook15}%
  \BibitemOpen
  \bibfield  {author} {\bibinfo {author} {\bibfnamefont {B.}~\bibnamefont
  {Zeng}}, \bibinfo {author} {\bibfnamefont {X.}~\bibnamefont {Chen}}, \bibinfo
  {author} {\bibfnamefont {D.-L.}\ \bibnamefont {Zhou}}, \ and\ \bibinfo
  {author} {\bibfnamefont {X.-G.}\ \bibnamefont {Wen}},\ }\href@noop {}
  {\bibfield  {journal} {\bibinfo  {journal} {arXiv:1508.02595}\ } (\bibinfo
  {year} {2015})}\BibitemShut {NoStop}%
\bibitem [{\citenamefont {Wen}\ \emph {et~al.}(2015)\citenamefont {Wen},
  \citenamefont {Chang},\ and\ \citenamefont {Ryu}}]{Ryu15}%
  \BibitemOpen
  \bibfield  {author} {\bibinfo {author} {\bibfnamefont {X.}~\bibnamefont
  {Wen}}, \bibinfo {author} {\bibfnamefont {P.-Y.}\ \bibnamefont {Chang}}, \
  and\ \bibinfo {author} {\bibfnamefont {S.}~\bibnamefont {Ryu}},\ }\href
  {\doibase 10.1103/PhysRevB.92.075109} {\bibfield  {journal} {\bibinfo
  {journal} {Phys. Rev. B}\ }\textbf {\bibinfo {volume} {92}},\ \bibinfo
  {pages} {075109} (\bibinfo {year} {2015})}\BibitemShut {NoStop}%
\bibitem [{\citenamefont {Islam}\ \emph {et~al.}(2015)\citenamefont {Islam},
  \citenamefont {Ma}, \citenamefont {Preiss}, \citenamefont {Tai},
  \citenamefont {Lukin}, \citenamefont {Rispoli},\ and\ \citenamefont
  {Greiner}}]{Greiner15}%
  \BibitemOpen
  \bibfield  {author} {\bibinfo {author} {\bibfnamefont {R.}~\bibnamefont
  {Islam}}, \bibinfo {author} {\bibfnamefont {R.}~\bibnamefont {Ma}}, \bibinfo
  {author} {\bibfnamefont {P.~M.}\ \bibnamefont {Preiss}}, \bibinfo {author}
  {\bibfnamefont {M.~E.}\ \bibnamefont {Tai}}, \bibinfo {author} {\bibfnamefont
  {A.}~\bibnamefont {Lukin}}, \bibinfo {author} {\bibfnamefont
  {M.}~\bibnamefont {Rispoli}}, \ and\ \bibinfo {author} {\bibfnamefont
  {M.}~\bibnamefont {Greiner}},\ }\href@noop {} {\bibfield  {journal} {\bibinfo
   {journal} {Nature}\ }\textbf {\bibinfo {volume} {528}},\ \bibinfo {pages}
  {77} (\bibinfo {year} {2015})}\BibitemShut {NoStop}%
\bibitem [{\citenamefont {Calabrese}\ and\ \citenamefont {Cardy}()}]{Cardy16}%
  \BibitemOpen
  \bibfield  {author} {\bibinfo {author} {\bibfnamefont {P.}~\bibnamefont
  {Calabrese}}\ and\ \bibinfo {author} {\bibfnamefont {J.}~\bibnamefont
  {Cardy}},\ }\href@noop {} {\bibinfo  {journal} {arXiv:1603.02889}\
  }\BibitemShut {NoStop}%
\bibitem [{\citenamefont {Calabrese}\ and\ \citenamefont
  {Cardy}(2005)}]{calabrese2005}%
  \BibitemOpen
\bibfield  {journal} {  }\bibfield  {author} {\bibinfo {author} {\bibfnamefont
  {P.}~\bibnamefont {Calabrese}}\ and\ \bibinfo {author} {\bibfnamefont
  {J.}~\bibnamefont {Cardy}},\ }\href@noop {} {\bibfield  {journal} {\bibinfo
  {journal} {Journal of Statistical Mechanics: Theory and Experiment}\ }\textbf
  {\bibinfo {volume} {2005}},\ \bibinfo {pages} {P04010} (\bibinfo {year}
  {2005})}\BibitemShut {NoStop}%
\bibitem [{\citenamefont {Calabrese}\ and\ \citenamefont
  {Cardy}(2007{\natexlab{b}})}]{calabrese2007ent}%
  \BibitemOpen
  \bibfield  {author} {\bibinfo {author} {\bibfnamefont {P.}~\bibnamefont
  {Calabrese}}\ and\ \bibinfo {author} {\bibfnamefont {J.}~\bibnamefont
  {Cardy}},\ }\href@noop {} {\bibfield  {journal} {\bibinfo  {journal} {Journal
  of Statistical Mechanics: Theory and Experiment}\ }\textbf {\bibinfo {volume}
  {2007}},\ \bibinfo {pages} {P10004} (\bibinfo {year}
  {2007}{\natexlab{b}})}\BibitemShut {NoStop}%
\bibitem [{\citenamefont {Igl\'oi}\ \emph {et~al.}(2012)\citenamefont
  {Igl\'oi}, \citenamefont {Szatm\'ari},\ and\ \citenamefont {Lin}}]{Igloi12}%
  \BibitemOpen
  \bibfield  {author} {\bibinfo {author} {\bibfnamefont {F.}~\bibnamefont
  {Igl\'oi}}, \bibinfo {author} {\bibfnamefont {Z.}~\bibnamefont {Szatm\'ari}},
  \ and\ \bibinfo {author} {\bibfnamefont {Y.-C.}\ \bibnamefont {Lin}},\ }\href
  {\doibase 10.1103/PhysRevB.85.094417} {\bibfield  {journal} {\bibinfo
  {journal} {Phys. Rev. B}\ }\textbf {\bibinfo {volume} {85}},\ \bibinfo
  {pages} {094417} (\bibinfo {year} {2012})}\BibitemShut {NoStop}%
\bibitem [{\citenamefont {Kim}\ and\ \citenamefont {Huse}(2013)}]{Huse13}%
  \BibitemOpen
  \bibfield  {author} {\bibinfo {author} {\bibfnamefont {H.}~\bibnamefont
  {Kim}}\ and\ \bibinfo {author} {\bibfnamefont {D.~A.}\ \bibnamefont {Huse}},\
  }\href {\doibase 10.1103/PhysRevLett.111.127205} {\bibfield  {journal}
  {\bibinfo  {journal} {Phys. Rev. Lett.}\ }\textbf {\bibinfo {volume} {111}},\
  \bibinfo {pages} {127205} (\bibinfo {year} {2013})}\BibitemShut {NoStop}%
\bibitem [{\citenamefont {Hartman}\ and\ \citenamefont
  {Maldacena}(2013)}]{hartman2013}%
  \BibitemOpen
  \bibfield  {author} {\bibinfo {author} {\bibfnamefont {T.}~\bibnamefont
  {Hartman}}\ and\ \bibinfo {author} {\bibfnamefont {J.}~\bibnamefont
  {Maldacena}},\ }\href@noop {} {\bibfield  {journal} {\bibinfo  {journal}
  {Journal of High Energy Physics}\ }\textbf {\bibinfo {volume} {135}},\
  \bibinfo {pages} {14} (\bibinfo {year} {2013})}\BibitemShut {NoStop}%
\bibitem [{\citenamefont {Chiocchetta}\ \emph {et~al.}(2015)\citenamefont
  {Chiocchetta}, \citenamefont {Tavora}, \citenamefont {Gambassi},\ and\
  \citenamefont {Mitra}}]{Chiocchetta2015}%
  \BibitemOpen
  \bibfield  {author} {\bibinfo {author} {\bibfnamefont {A.}~\bibnamefont
  {Chiocchetta}}, \bibinfo {author} {\bibfnamefont {M.}~\bibnamefont {Tavora}},
  \bibinfo {author} {\bibfnamefont {A.}~\bibnamefont {Gambassi}}, \ and\
  \bibinfo {author} {\bibfnamefont {A.}~\bibnamefont {Mitra}},\ }\href
  {\doibase 10.1103/PhysRevB.91.220302} {\bibfield  {journal} {\bibinfo
  {journal} {Phys. Rev. B}\ }\textbf {\bibinfo {volume} {91}},\ \bibinfo
  {pages} {220302} (\bibinfo {year} {2015})}\BibitemShut {NoStop}%
\bibitem [{\citenamefont {Maraga}\ \emph {et~al.}(2015)\citenamefont {Maraga},
  \citenamefont {Chiocchetta}, \citenamefont {Mitra},\ and\ \citenamefont
  {Gambassi}}]{Maraga2015}%
  \BibitemOpen
  \bibfield  {author} {\bibinfo {author} {\bibfnamefont {A.}~\bibnamefont
  {Maraga}}, \bibinfo {author} {\bibfnamefont {A.}~\bibnamefont {Chiocchetta}},
  \bibinfo {author} {\bibfnamefont {A.}~\bibnamefont {Mitra}}, \ and\ \bibinfo
  {author} {\bibfnamefont {A.}~\bibnamefont {Gambassi}},\ }\href {\doibase
  10.1103/PhysRevE.92.042151} {\bibfield  {journal} {\bibinfo  {journal} {Phys.
  Rev. E}\ }\textbf {\bibinfo {volume} {92}},\ \bibinfo {pages} {042151}
  (\bibinfo {year} {2015})}\BibitemShut {NoStop}%
\bibitem [{\citenamefont {Chiocchetta}\ \emph {et~al.}(2016)\citenamefont
  {Chiocchetta}, \citenamefont {Tavora}, \citenamefont {Gambassi},\ and\
  \citenamefont {Mitra}}]{Chiocchetta16}%
  \BibitemOpen
  \bibfield  {author} {\bibinfo {author} {\bibfnamefont {A.}~\bibnamefont
  {Chiocchetta}}, \bibinfo {author} {\bibfnamefont {M.}~\bibnamefont {Tavora}},
  \bibinfo {author} {\bibfnamefont {A.}~\bibnamefont {Gambassi}}, \ and\
  \bibinfo {author} {\bibfnamefont {A.}~\bibnamefont {Mitra}},\ }\href@noop {}
  {\bibfield  {journal} {\bibinfo  {journal} {arXiv:1604.04614}\ } (\bibinfo
  {year} {2016})}\BibitemShut {NoStop}%
\bibitem [{\citenamefont {Chandran}\ \emph {et~al.}(2013)\citenamefont
  {Chandran}, \citenamefont {Nanduri}, \citenamefont {Gubser},\ and\
  \citenamefont {Sondhi}}]{Sondhi2013}%
  \BibitemOpen
  \bibfield  {author} {\bibinfo {author} {\bibfnamefont {A.}~\bibnamefont
  {Chandran}}, \bibinfo {author} {\bibfnamefont {A.}~\bibnamefont {Nanduri}},
  \bibinfo {author} {\bibfnamefont {S.~S.}\ \bibnamefont {Gubser}}, \ and\
  \bibinfo {author} {\bibfnamefont {S.~L.}\ \bibnamefont {Sondhi}},\ }\href
  {\doibase 10.1103/PhysRevB.88.024306} {\bibfield  {journal} {\bibinfo
  {journal} {Phys. Rev. B}\ }\textbf {\bibinfo {volume} {88}},\ \bibinfo
  {pages} {024306} (\bibinfo {year} {2013})}\BibitemShut {NoStop}%
\bibitem [{\citenamefont {Janssen}\ \emph {et~al.}(1989)\citenamefont
  {Janssen}, \citenamefont {Schaub},\ and\ \citenamefont
  {Schmittmann}}]{Janssen1988}%
  \BibitemOpen
  \bibfield  {author} {\bibinfo {author} {\bibfnamefont {H.}~\bibnamefont
  {Janssen}}, \bibinfo {author} {\bibfnamefont {B.}~\bibnamefont {Schaub}}, \
  and\ \bibinfo {author} {\bibfnamefont {B.}~\bibnamefont {Schmittmann}},\
  }\href {\doibase 10.1007/BF01319383} {\bibfield  {journal} {\bibinfo
  {journal} {Z. Phys. B}\ }\textbf {\bibinfo {volume} {73}},\ \bibinfo {pages}
  {539} (\bibinfo {year} {1989})}\BibitemShut {NoStop}%
\bibitem [{\citenamefont {Huse}(1989)}]{Huse89}%
  \BibitemOpen
  \bibfield  {author} {\bibinfo {author} {\bibfnamefont {D.~A.}\ \bibnamefont
  {Huse}},\ }\href {\doibase 10.1103/PhysRevB.40.304} {\bibfield  {journal}
  {\bibinfo  {journal} {Phys. Rev. B}\ }\textbf {\bibinfo {volume} {40}},\
  \bibinfo {pages} {304} (\bibinfo {year} {1989})}\BibitemShut {NoStop}%
\bibitem [{\citenamefont {Calabrese}\ and\ \citenamefont
  {Gambassi}(2005)}]{Gambassi2005}%
  \BibitemOpen
  \bibfield  {author} {\bibinfo {author} {\bibfnamefont {P.}~\bibnamefont
  {Calabrese}}\ and\ \bibinfo {author} {\bibfnamefont {A.}~\bibnamefont
  {Gambassi}},\ }\href {http://stacks.iop.org/0305-4470/38/i=18/a=R01}
  {\bibfield  {journal} {\bibinfo  {journal} {J. Phys. A: Math. Gen.}\ }\textbf
  {\bibinfo {volume} {38}},\ \bibinfo {pages} {R133} (\bibinfo {year}
  {2005})}\BibitemShut {NoStop}%
\bibitem [{\citenamefont {Gagel}\ \emph {et~al.}(2014)\citenamefont {Gagel},
  \citenamefont {Orth},\ and\ \citenamefont {Schmalian}}]{Gagel2014}%
  \BibitemOpen
  \bibfield  {author} {\bibinfo {author} {\bibfnamefont {P.}~\bibnamefont
  {Gagel}}, \bibinfo {author} {\bibfnamefont {P.~P.}\ \bibnamefont {Orth}}, \
  and\ \bibinfo {author} {\bibfnamefont {J.}~\bibnamefont {Schmalian}},\ }\href
  {\doibase 10.1103/PhysRevLett.113.220401} {\bibfield  {journal} {\bibinfo
  {journal} {Phys. Rev. Lett.}\ }\textbf {\bibinfo {volume} {113}},\ \bibinfo
  {pages} {220401} (\bibinfo {year} {2014})}\BibitemShut {NoStop}%
\bibitem [{\citenamefont {Kibble}(1976)}]{Kibble76}%
  \BibitemOpen
  \bibfield  {author} {\bibinfo {author} {\bibfnamefont {T.~W.~B.}\
  \bibnamefont {Kibble}},\ }\href {http://stacks.iop.org/0305-4470/9/i=8/a=029}
  {\bibfield  {journal} {\bibinfo  {journal} {Journal of Physics A:
  Mathematical and General}\ }\textbf {\bibinfo {volume} {9}},\ \bibinfo
  {pages} {1387} (\bibinfo {year} {1976})}\BibitemShut {NoStop}%
\bibitem [{\citenamefont {Zurek}(1985)}]{Zurek85}%
  \BibitemOpen
  \bibfield  {author} {\bibinfo {author} {\bibfnamefont {W.~H.}\ \bibnamefont
  {Zurek}},\ }\href {\doibase 10.1038/317505a0} {\bibfield  {journal} {\bibinfo
   {journal} {Nature}\ }\textbf {\bibinfo {volume} {317}},\ \bibinfo {pages}
  {505} (\bibinfo {year} {1985})}\BibitemShut {NoStop}%
\bibitem [{\citenamefont {Sciolla}\ and\ \citenamefont
  {Biroli}(2013)}]{Sciolla2013}%
  \BibitemOpen
  \bibfield  {author} {\bibinfo {author} {\bibfnamefont {B.}~\bibnamefont
  {Sciolla}}\ and\ \bibinfo {author} {\bibfnamefont {G.}~\bibnamefont
  {Biroli}},\ }\href {\doibase 10.1103/PhysRevB.88.201110} {\bibfield
  {journal} {\bibinfo  {journal} {Phys. Rev. B}\ }\textbf {\bibinfo {volume}
  {88}},\ \bibinfo {pages} {201110} (\bibinfo {year} {2013})}\BibitemShut
  {NoStop}%
\bibitem [{\citenamefont {Gambassi}\ and\ \citenamefont
  {Calabrese}(2011)}]{Gambassi11}%
  \BibitemOpen
  \bibfield  {author} {\bibinfo {author} {\bibfnamefont {A.}~\bibnamefont
  {Gambassi}}\ and\ \bibinfo {author} {\bibfnamefont {P.}~\bibnamefont
  {Calabrese}},\ }\href {http://stacks.iop.org/0295-5075/95/i=6/a=66007}
  {\bibfield  {journal} {\bibinfo  {journal} {EPL (Europhysics Letters)}\
  }\textbf {\bibinfo {volume} {95}},\ \bibinfo {pages} {66007} (\bibinfo {year}
  {2011})}\BibitemShut {NoStop}%
\bibitem [{\citenamefont {Smacchia}\ \emph {et~al.}(2015)\citenamefont
  {Smacchia}, \citenamefont {Knap}, \citenamefont {Demler},\ and\ \citenamefont
  {Silva}}]{Smacchia2015}%
  \BibitemOpen
  \bibfield  {author} {\bibinfo {author} {\bibfnamefont {P.}~\bibnamefont
  {Smacchia}}, \bibinfo {author} {\bibfnamefont {M.}~\bibnamefont {Knap}},
  \bibinfo {author} {\bibfnamefont {E.}~\bibnamefont {Demler}}, \ and\ \bibinfo
  {author} {\bibfnamefont {A.}~\bibnamefont {Silva}},\ }\href {\doibase
  10.1103/PhysRevB.91.205136} {\bibfield  {journal} {\bibinfo  {journal} {Phys.
  Rev. B}\ }\textbf {\bibinfo {volume} {91}},\ \bibinfo {pages} {205136}
  (\bibinfo {year} {2015})}\BibitemShut {NoStop}%
\bibitem [{\citenamefont {Peschel}\ and\ \citenamefont
  {Eisler}(2009)}]{Eisler2009}%
  \BibitemOpen
  \bibfield  {author} {\bibinfo {author} {\bibfnamefont {I.}~\bibnamefont
  {Peschel}}\ and\ \bibinfo {author} {\bibfnamefont {V.}~\bibnamefont
  {Eisler}},\ }\href {http://stacks.iop.org/1751-8121/42/i=50/a=504003}
  {\bibfield  {journal} {\bibinfo  {journal} {Journal of Physics A:
  Mathematical and Theoretical}\ }\textbf {\bibinfo {volume} {42}},\ \bibinfo
  {pages} {504003} (\bibinfo {year} {2009})}\BibitemShut {NoStop}%
\bibitem [{\citenamefont {Li}\ and\ \citenamefont {Haldane}(2008)}]{Haldane08}%
  \BibitemOpen
  \bibfield  {author} {\bibinfo {author} {\bibfnamefont {H.}~\bibnamefont
  {Li}}\ and\ \bibinfo {author} {\bibfnamefont {F.~D.~M.}\ \bibnamefont
  {Haldane}},\ }\href {\doibase 10.1103/PhysRevLett.101.010504} {\bibfield
  {journal} {\bibinfo  {journal} {Phys. Rev. Lett.}\ }\textbf {\bibinfo
  {volume} {101}},\ \bibinfo {pages} {010504} (\bibinfo {year}
  {2008})}\BibitemShut {NoStop}%
\bibitem [{\citenamefont {Berges}\ and\ \citenamefont
  {Serreau}()}]{Berges2003}%
  \BibitemOpen
  \bibfield  {author} {\bibinfo {author} {\bibfnamefont {J.}~\bibnamefont
  {Berges}}\ and\ \bibinfo {author} {\bibfnamefont {J.}~\bibnamefont
  {Serreau}},\ }\href@noop {} {\bibinfo  {journal} {arXiv:hep-ph/0302210,
  hep-ph/0410330}\ }\BibitemShut {NoStop}%
\end{thebibliography}

%

\appendix
\section{Energy density of quench states}

We work in terms of the function $f_k$. For a free field the energy is given by
\begin{equation}
E = \sum_k\left[|f_k|^2\varepsilon_k^2 + |\partial_t f_k|^2\right]. \label{eq:energy}
\end{equation}
In the free quench the function $f$ is given by,
\begin{equation}
f_k(t) = \frac{1}{\sqrt{2\omega_0}} \cos \varepsilon_k t   + i\sqrt{\frac{\omega_0}{2}}\frac{\sin \varepsilon_k t}{\varepsilon_k}.
\end{equation}
Plugging into Eq.~\eqref{eq:energy} we obtain
\begin{equation}
E= \frac{1}{2} \left[ \frac{\langle \varepsilon_k^2\rangle}{\omega_0} + \omega_0\right].
\end{equation}
where $\langle \cdot \rangle$ denotes average over $k$. For the nearest neighbor hopping dispersion
\begin{equation}
{\varepsilon}_k = \sqrt{\sum_{i=1\ldots d} \left(1 - \cos(k_i)\right)},
\end{equation}
we have $ \langle \varepsilon_k^2\rangle = 3$.
For the interacting case we have,
\begin{eqnarray}
&&f_k = \sqrt{t}\biggl[i\left(\frac{\varepsilon_k}{\omega_0}\right)^{-\alpha} j_\alpha \left(\varepsilon_k t\right) \nonumber\\
&&- \frac{\pi }{4 \sin \pi \alpha} \left(\frac{\varepsilon_k}{\omega_0}\right)^\alpha  j_{-\alpha} \left(\varepsilon_k t\right) \biggr],
\end{eqnarray}
where we have set
\begin{equation}
\kappa_+ = i\sqrt{2}
\qquad
\kappa_- = -\frac{\pi}{2\sqrt{2}\sin \pi \alpha}.
\end{equation}
Now in the limit $t\rightarrow\infty$, the effective mass $r_{\rm eff}(t)$ goes to zero so that the Hamiltonian becomes that of the free case. $f_k$
after taking the long time limit of the Bessel functions, reduces to,
\begin{eqnarray}
f_k =&&i\sqrt{\frac{2}{\pi \varepsilon_k}}
\left(\frac{\varepsilon_k}{\omega_0}\right)^{-\alpha}
\!\!\sin\left(
	\varepsilon_k t - \frac{\pi \alpha}{2} + \frac{\pi}{4}
\right)\nonumber\\
&&-
\frac{\pi }{4 \sin \pi \alpha}
\sqrt{\frac{2}{\pi \varepsilon_k}}
\left(\frac{\varepsilon_k}{\omega_0}\right)^{\alpha}
\!\!\sin\left(\varepsilon_k t + \frac{\pi \alpha}{2}+ \frac{\pi}{4}\right).
\end{eqnarray}

Plugging this into the energy functional one obtains
\begin{align}
E &= \Bigg\langle
\varepsilon_k^2\left[
	\sqrt{\frac{2}{\pi \varepsilon_k}}
	\left(\frac{\varepsilon_k}{\omega_0}\right)^{-\alpha}
\right]^2\nonumber
\\&\qquad+
\varepsilon_k^2\left[
	\frac{\pi }{4 \sin \pi \alpha}
	\sqrt{\frac{2}{\pi \varepsilon_k}}
	\left(\frac{\varepsilon_k}{\omega_0}\right)^{\alpha}
\right]^2
\Bigg\rangle
\\
&= \frac{2\omega^{2\alpha}_0}{\pi}
\left\langle
	\varepsilon_k^{-2\alpha +1}
\right\rangle
+
\frac{\pi}{8\omega^{2\alpha}_0\sin^2 \pi\alpha}
\left\langle
	\varepsilon_k^{2\alpha +1}.
\right\rangle
\end{align}

(This reduces to the free case with $\alpha = 1/2$ after the redefinition $\omega_0 \rightarrow \pi\omega_0/4$). For the n.n. dispersion and the physical case of $\alpha =1/4$ mathematica evaluates
\begin{align}
\left\langle
	{\varepsilon}_k^{1/2}
\right\rangle
&\approx
1.533.
\\
\left\langle
	{\varepsilon}_k^{3/2}
\right\rangle
&\approx
3.764.
\end{align}

We obtained that for the free state to have equal energy to the interacting state with $\omega_0 = 100$, the free state should have  $\omega_0 \approx 19.98$.
For the altered dispersion
\begin{equation}
\tilde{\varepsilon}_k
= \sqrt{
	\sum_{i=1\ldots d} \left(1 - \cos(k_i)\right)
	+
	\left[
		\sum_{i=1\ldots d} \left(1 - \cos(k_i)\right)
	\right]^2
},
\end{equation}
and for the physical case of $\alpha =1/4$, the integrals evaluate to
\begin{align}
\left\langle
	\tilde{\varepsilon}_k^{1/2}
\right\rangle
&\approx
1.931.
\\
\left\langle
	\tilde{\varepsilon}_k^{3/2}
\right\rangle
&\approx
7.888.
\\
\left\langle
	\tilde{\varepsilon}_k^{2}
\right\rangle
&=
33/2.
\end{align}
\clearpage

\end{document}